\documentstyle[aps,preprint,pre,epsf,epsfig]{revtex}
\begin{document}         
%
%
%

\title{The Statistical Physics of Regular Low-Density Parity-Check Error-Correcting Codes}
\author{Tatsuto Murayama$^{1}, $Yoshiyuki~Kabashima$^{1}$, David~Saad$^{2}$
and Renato~Vicente$^{2}$}
\address{$^{1}$ Department of Computational Intelligence and Systems Science,
Tokyo Institute of Technology, Yokohama 2268502, Japan.  \\
$^{2}$The Neural Computing Research Group, Aston
University, Birmingham B4 7ET, UK.}
%
\maketitle
\begin{abstract}      
A variation of Gallager error-correcting codes is investigated using
statistical mechanics.  In codes of this type, a given message is
encoded into a codeword which comprises Boolean sums of message bits
selected by two randomly constructed sparse matrices.  The similarity
of these codes to Ising spin systems with random interaction makes it
possible to assess their typical performance by analytical methods
developed in the study of disordered systems.  The typical case
solutions obtained via the replica method are consistent with those
obtained in simulations using belief propagation (BP) decoding. We
discuss the practical implications of the results obtained and suggest
a computationally efficient construction for one of the more practical
configurations. 

\end{abstract}
%

\newcommand{\mltau}{\mbox{\large\boldmath{$\tau$}}}
\newcommand{\mtau}{\mbox{\boldmath{$\tau$}}}
\newcommand{\mxi}{\mbox{\boldmath{$\xi$}}}
\newcommand{\mzeta}{\mbox{\boldmath{$\zeta$}}}
\newcommand{\mzzeta}{\mbox{\boldmath{$\zzeta$}}}
\newcommand{\tr}{{\!{\mbox{\tiny T}}}}
\newcommand{\rd}{{\mathrm d}}
\newcommand{\e}{{\mathrm e}}
\newcommand{\m}{{\mathrm m}}
\newcommand{\mS}{{\mbox{{\boldmath $S$}}}}
\newcommand{\G}{\mbox{${\bf G}$}}
\newcommand{\eg}{\mbox{\large $\epsilon$}_{g}}
\newcommand{\er}{\mbox{\large $\epsilon$}}
\newcommand{\J}{{\mbox{{\boldmath $J$}}}}
\newcommand{\bm}{{\mbox{{\boldmath $m$}}}}
\newcommand{\mz}{{\mbox{{\boldmath $z$}}}}
\newcommand{\z}{{\mbox{{\boldmath $z$}}}}
\newcommand{\mt}{{\mbox{{\boldmath $t$}}}}
\newcommand{\mcJ}{{\mbox{{\boldmath ${\cal J}$}}}}
\newcommand{\cD}{{\cal D}}
\newcommand{\cJ}{{\cal J}}

\section{Introduction}
Error-correcting codes are commonly used for reliable data
transmission through noisy media, especially in the case of memoryless
communication where corrupted messages cannot be repeatedly sent.  These
techniques play an important role in a wide range of applications from
memory devices to deep space explorations, and are expected to become
even more important due to the rapid development in mobile phones and
satellite-based communication.

In a general scenario, the sender encodes an $N$ dimensional Boolean
message vector $\mxi$, where $\xi_{i}\in (0,1), \ \forall i$, to an
$M(>N)$ dimensional Boolean codeword $\mz_0$, which is then being
transmitted through a noisy communication channel. Noise corruption
during transmission can be modelled by the noise vector $\mzeta$,
where corrupted bits are marked by the value 1 and all other bits are
zero, such that the received corrupted codeword takes the form
$\mz=\mz_0+\mzeta \ \mbox{(mod 2)}$.  The received corrupted message
is then decoded by the receiver for retrieving the original message
$\mxi$.

The error-correcting ability comes at the expense of information
redundancy.  Shannon showed in his seminal work\cite{Shannon} that
error-free communication is theoretically possible if the code rate,
representing the fraction of informative bits in the transmitted
codeword, is below the channel capacity; in the case of unbiased
messages transmitted through a Binary Symmetric Channel (BSC), which
we will focus on here, $R=N/M$ satisfies
\begin{equation}
R < 1+p\log_2 p + (1-p)\log_2(1-p) \ . 
\label{eq:shannon_capacity}
\end{equation}
The expression on the right is termed {\em Shannon's bound}.  However,
Shannon's derivation is non-constructive and the quest for codes which
saturate Eq.(\ref{eq:shannon_capacity}) has been one of the central
topics of information theory ever since.

In this paper we examine the efficiency and limitations of
Gallager-type error-correcting code\cite{Gallager,MN}, which attracted
much interest recently among researchers in this field.  This code was
discovered almost forty years ago by Gallager\cite{Gallager} but was
abandoned shortly after its invention due to the computational
limitations of the time.  Since their recent rediscovery by MacKay and
Neal\cite{MN}, different variations of Gallager-type codes have been
developed \cite{Kanter_Saad,Luby,MacKay,Richardson} attempting to get
as close as possible to saturating Shannon's bound.

In this paper we will examine the typical properties of a family of
codes based on one variation, the MN code \cite{MN}, using the
established methods of statistical
physics\cite{us_TAP_sourlas,us_replica_sourlas,us_PRL,Vicente}, to
provide a theoretical study based on the typical performance of codes
rather on the worst case analysis.

This paper is organised as follows: In the next two sections, we
introduce Gallager-type error-correcting codes in detail and link them
to the statistical mechanics framework.  We then examine the
equilibrium properties of various members of this family of codes
using the replica method (section IV) and compare the bit-error rate
below criticality.  In section V, we examine the relation between
Belief-Propagation (BP) decoding and the Thouless-Anderson-Palmer
(TAP) approach to diluted spin systems; we then use it for comparing
the analytical results obtained via the replica method to those
obtained from simulations using BP decoding.  In section VI we show a
computationally efficient construction for one of the more practical
constructions. Finally, we present conclusions for the current work
and suggest future research directions.

\section{Gallager-type error-correcting codes}
There are several variations in Gallager-type error-correcting codes. 
The one discussed in this paper is termed the {\em MN} code,
recently introduced by MacKay and Neal\cite{MN}. 
In these codes, a Boolean message $\mxi$ is encoded into a codeword $\mz_0$ using
two randomly constructed Boolean sparse matrices $C_s$ and $C_n$, which 
are characterised in the following manner. 

The rectangular sparse matrix $C_s$ is of dimensionality $M \times N$,
having randomly chosen $K$ non-zero unit elements per row and $C$ per
column.  The matrix $C_n$ is an $M \times M$ (mod 2)-invertible matrix
having randomly chosen $L$ non-zero elements per row and column.
These matrices are shared by the sender and the receiver.

Using these matrices, one can encode a message $\mxi$ 
into a codeword $\z_0$ in the following manner
\begin{equation}
\mz_0=C_n^{-1} C_s \mxi \quad \mbox{(mod 2)},
\end{equation}
which is then transmitted via a noisy channel. Note that all matrix
and vector components are Boolean $(0,1)$, and all summations are
carried out in this field.  For simplicity, the noise process is
modelled hereafter by a binary symmetric channel (BSC), where each bit
is independently flipped with probability $p$.  Extending the code
presented here to other types of noise is straightforward.

During transmission, a noise vector $\mzeta$ is added to
$\mz_0$ and a corrupted codeword $\mz=\mz_0+\mzeta$ (mod 2) is
received at the other end of the channel.  Decoding is then carried
out by taking the product of the matrix $C_n$ and the received
codeword $\mz$, which results in $C_s \mxi +C_n \mzeta =C_n \z \equiv
\bbox{J}$.  The equation
\begin{equation}
\label{eq:decoding}
C_s \mS +C_n \mtau =  \J \quad \mbox{(mod 2)},
\end{equation}
is solved via the iterative methods of belief propagation
(BP)\cite{Frey,Pearl} to obtain the most probable Boolean vectors
$\mS$ and $\mtau$.  BP methods in this context have recently been
shown to be identical to a TAP\cite{TAP} based solution of a similar
physical system\cite{us_TAP_sourlas}.

\section{A Statistical Mechanics Perspective}
Sourlas was the first to point out that error-correcting codes of this
type have a similarity to Ising spin systems in statistical
physics\cite{Sourlas_Nature}; he demonstrated this using a simple
version of the same nature.  His
work, that focused on extensively connected systems, was recently
extended to finitely connected systems\cite{us_replica_sourlas,Vicente}.  We
follow a similar approach in the current investigation; preliminary
results have already been presented in\cite{us_PRL}.

To facilitate the statistical physics analysis, we first employ the
binary representation $(\pm 1)$ of the dynamical variables $\mS$ and
$\mtau$ and of the check vector $\J$ rather than the Boolean one
$(0,1)$. The $\mu$-th component of Eq.(\ref{eq:decoding}) is then
rewritten as
\begin{equation}
\prod_{i \in {\cal L}_s(\mu)}S_i 
\prod_{j \in {\cal L}_n(\mu)}\tau_j=J_\mu,
\label{eq:Ising_condition}
\end{equation}
where ${\cal L}_s(\mu)$ and ${\cal L}_n(\mu)$ are the sets of all
indices of non-zero elements in row $\mu$ of the sparse matrices $C_s$
and $C_n$, respectively.  The check $\mu$ is given by message $\mxi$
and noise $\mzeta$ as $J_\mu = \prod_{i \in {\cal L}_s(\mu)}\xi_i
\prod_{j \in {\cal L}_n(\mu)}\zeta_j$; it should be emphasised that
the message vector $\mxi$ and the noise vector $\mzeta$ themselves are
not known to the receiver.

An interesting link can now be formulated between the Bayesian
framework of {\em MN} codes and Ising spin systems.  Rewriting
Kronecker's delta for binary variables $x$ and $y$ as
$\delta[x;y]=(1+xy)/2= \lim_{\beta \to \infty} \exp ( -\beta
\delta[-1;xy]) $, one may argue that, using it as a likelihood, equation
(\ref{eq:Ising_condition}) gives rise to the conditional probability
of the check $\J$ for given $\mS$, $\mtau$, $C_s$ and $C_n$
\begin{eqnarray}
{\cal P}(\bbox{J}|\bbox{S},\bbox{\tau},C_s,C_n) = \lim_{\beta \to
\infty} \exp \left ( -\beta \sum_{\mu=1}^M \delta [-1;J_{\mu} \prod_{i
\in {\cal L}_s(\mu)} S_i \prod_{j \in {\cal L}_n(\mu)} \tau_j] \right
).
\label{eq:cond_prob}
\end{eqnarray}
Prior knowledge about possibly biased message and noise is
represented by the prior distributions
\begin{eqnarray}
{\cal P}_s(\bbox{S})
 ={ \exp    \left(
      F_s \sum_{i=1}^N S_i
    \right) \over \left ( 2 \cosh F_s \right )^N},\quad
{\cal P}_n(\bbox{\tau}) = { \exp \left(
      F_n \sum_{j=1}^M \tau_j
    \right) \over  \left ( 2 \cosh F_n \right )^M}, 
   \label{eq:prior}
\end{eqnarray}
respectively. Non-zero field $F_s$ is introduced for biased message
and $F_n$ is determined by flip rate $p$ of channel noise as
$F_n=(1/2) \ln \left ((1-p)/p \right )$.  Using equations
(\ref{eq:cond_prob}) and (\ref{eq:prior}), the posterior distribution
of $\mS$ and $\mtau$ for given check $\J$ and matrices $C_s$ and $C_n$
is of the form
\begin{eqnarray}
{\cal P} \left (\mS,\mtau | \J, C_s, C_n \right ) &=& \frac{
{\cal P}(\bbox{J}|\bbox{S},\bbox{\tau},  C_s, C_n ) {\cal P}_s(\bbox{S}) 
{\cal P}_n(\bbox{\tau}) }
{{\cal P} \left (\J| C_s,C_n \right )} \cr
&=& \lim_{\beta \to \infty} 
\frac{\exp \left ( -\beta {\cal H} (\mS,\mtau|\mcJ,\cD) \right )}
{{\cal Z} (\mcJ,\cD) }, 
\label{eq:posterior}
\end{eqnarray}
where ${\cal P}(\J|C_s,C_n)=\sum_{\{\mS,\mtau \}} {\cal P}(\bbox{J}|\bbox{S},\bbox{\tau},C_s,C_n) {\cal P}_s(\bbox{S}) 
{\cal P}_n(\bbox{\tau})$, 
\begin{eqnarray}
{\cal H}(\mS,\mtau|\mcJ,\cD) &=& \sum_{\mu=1}^M 
\delta[-1;J_{\mu} \!\!\! \prod_{i \in {\cal L}_s(\mu)} 
\!\!\! S_i \!\!\!
\prod_{j \in {\cal L}_n(\mu)} \!\!\! \tau_j] - 
\frac{F_s}{\beta} \sum_{i=1}^N S_i - \frac{F_n}{\beta}\sum_{j=1}^M \tau_j \cr 
&=& \!\!\!\!\!\!\!\! \sum_{<i_1,..,i_K;j_1,..,j_L>}
 \mbox{\hspace*{-5mm}} \!\!\!\! \cD_{<i_1,..,i_K;j_1,..,j_L>} \delta
 \biggl[-1;\cJ_{<i_1,..,i_K;j_1,..,j_L>} 
  S_{i_1}\ldots S_{i_K} \tau_{j_1}\ldots\tau_{j_L}
 \biggr] \cr
&\phantom{=}& 
-\frac{F_s}{\beta} \sum_{i=1}^N S_i - \frac{F_n}{\beta}\sum_{j=1}^M \tau_j \ ,
\label{eq:hamiltonian}
\end{eqnarray}
and 
\begin{eqnarray}
{\cal Z}(\mcJ,\cD)
&= &\lim_{\beta \to \infty} \sum_{\{ \bbox{S},\bbox{\tau} \}} 
\exp \left ( -\beta {\cal H} (\mS,\mtau|\mcJ,\cD ) \right ) \cr
&=& \sum_{\{\bbox{S},\bbox{\tau}\}}
 \prod_{<i_1,..,i_K;j_1,..,j_L>}
 \mbox{\hspace*{-5mm}} \left[
      1+\frac{1}{2}{\cal D}_{\langle i_1,\cdots ,i_K;j_1,\cdots,j_L \rangle}
      \left(\cJ_{<i_1,..,i_K;j_1,..,j_L>} 
  S_{i_1}\ldots S_{i_K} \tau_{j_1}\ldots\tau_{j_L} 
-1 \right) \right] \cr
&\times&  \exp    \left(
      F_s \sum_{i=1}^N S_i
      +
      F_n \sum_{j=1}^M \tau_j
    \right). 
\label{eq:partition_function}
\end{eqnarray}
The final form of posterior distribution (\ref{eq:posterior}) implies
that the {\em MN} code is identical to an Ising spin system defined by
the Hamiltonian (\ref{eq:hamiltonian}) in the zero temperature limit
$T=\beta^{-1} \to 0$.  In equations (\ref{eq:hamiltonian}) and
(\ref{eq:partition_function}), we introduced the sparse connectivity
tensor $\cD_{<i_1,..,j_L>}$ which takes the value 1 if the
corresponding indices of both message and noise are chosen (i.e., if
all corresponding indices of the matrices $C_s$ and $C_n$ are 1) and 0
otherwise, and coupling $\cJ_{<i_1,..,i_K;j_1,..,j_L>}= \xi_{i_{1}}
\xi_{i_{2}} \ldots \xi_{i_{K}} \zeta_{j_{1}} \zeta_{j_{2}} \ldots
\zeta_{j_{L}}$. These come to isolate the disorder in choosing the
matrix connections, embedded in $\cD_{<i_1,..,j_L>}$, and to simplify
the notation.

The posterior distribution (\ref{eq:posterior}) can be used for decoding. 
One can show that expectation of the overlap between 
original message $\mxi$ and retrieved one $\hat{\mxi}$ 
\begin{equation}
m=\frac{1}{N} \sum_{i=1}^N \xi_i \hat{\xi}_i, 
\label{eq:overlap}
\end{equation}
is maximised by setting $\hat{\mxi}$ to its Bayes-optimal 
estimator\cite{Iba,Nishimori,Rujan,Sourlas_EPL}
\begin{equation}
\hat{\xi}^B_i=\mbox{sign}(m_i^S), \quad 
m_i^S = \sum_{\{\mS,\mtau \}} S_i \ {\cal P} (\mS,\mtau | \J, C_s,C_n). 
\protect\label{eq:bayesoptimal}
\end{equation}

It is worth while noting that this optimal decoding is realized at
{\em zero temperature} rather than at a {\em finite temperature} as in
\cite{Nishimori,Rujan,Sourlas_EPL}.  The reason is that the true
likelihood term (\ref{eq:cond_prob}) corresponds to the {\em ground
state} of the first term of the Hamiltonian (\ref{eq:hamiltonian}) due
to the existence of more degrees of freedom, in the form of the
dynamical variables $\mtau$, which do not appear in other systems.
Introducing the additional variables $\mtau$, the degrees of freedom
in the spin system increase from $N$ to $N+M$, while the number of
constraints from the checks $\J$ remains $M$.  This implies that in
spite of the existence of quenched disorder caused by $\mcJ$ and
$\cD$, the system is free from frustration even in the low temperature
limit, which is useful for practical decoding using local search
algorithms.  The last two terms in Eq.(\ref{eq:hamiltonian}) scale
with $\beta$ remain finite even in the zero temperature limit $\beta
\to \infty$ representing the true prior distributions, which dominates
the statistical properties of the system, while the first term
vanishes to satisfy the parity check condition
(\ref{eq:Ising_condition}).

\section{Equilibrium Properties: The Replica Method}
As we use the methods of statistical mechanics, we concentrate on the
case of long messages, in the of limit of $N, M \to \infty$ while
keeping code rate $R=N/M=K/C$ finite.  This limit is quite reasonable
for this particular problem since Gallager-type codes are usually
used in the transmission of long ($10^{4}\!-\!10^{5}$) messages, where
finite size corrections are likely to be negligible.

Since the first part of the Hamiltonian (\ref{eq:hamiltonian}) is
invariant under the gauge transformation $S_i \to \xi_i S_i$, $\tau_j
\to \zeta_j \tau_j$ and ${\cJ}_{\left \langle i_1,\ldots,j_L \right
\rangle } \to 1$, it is useful to decouple the correlation between the
vectors $\mS$, $\mtau$ and $\mxi$, $\mzeta$.  Rewriting the
Hamiltonian using this gauge, one obtains a similar expression to
Eq.(\ref{eq:hamiltonian}) apart from the second terms which become
$F_s/\beta \sum_{i=1} \xi_i S_i$ and $F_n/\beta \sum_{j=1} \zeta_j
\tau_j$.

Due to the existence of several types of quenched disorder in the system, 
it is natural to resort to replica 
method for investigating the typical properties in 
equilibrium.  More specifically, we calculate expectation values of 
$n$-th power of partition function (\ref{eq:partition_function}) 
with respect to the quenched variables $\mxi$, $\mzeta$ and $\cD$ 
and take the limit $n \to 0$. 

Carrying out the calculation in the zero temperature limit $\beta \to
\infty$ gives rise to a set of order parameters
\begin{eqnarray}
\label{eq:order_parameters}
q_{\alpha, \beta,.., \gamma} = \left\langle \frac{1}{N} \sum_{i=1}^{N}
Z_{i} \ S_{i}^{\alpha} \ S_{i}^{\beta},..,S_{i}^{\gamma}
\right\rangle_{\beta\rightarrow\infty}, \quad r_{\alpha, \beta,..,
\gamma} = \left\langle \frac{1}{M} \sum_{i=1}^{M} Y_{j} \
\tau_{j}^{\alpha} \ \tau_{j}^{\beta},.., \tau_{j}^{\gamma}
\right\rangle_{\beta\rightarrow\infty}
\end{eqnarray}
where $\alpha$, $\beta,..$ represent replica indices, and 
the variables $Z_{i}$ and $ Y_{j}$ come from enforcing the restriction 
of $C$ and $L$ connections per index,
respectively\cite{us_replica_sourlas,Wong}:
\begin{equation}
\label{eq:delta}
\delta \left( \sum_{\left\langle i_{2},..,i_{K}
\right\rangle} \!\!\!\!\!\! \cD_{<i, i_{2},..,j_L>} -
C \right) = \oint_{0}^{2 \pi} \frac{d Z}{2 \pi} \ 
Z^{\sum_{\left\langle i_{2},.., i_{K} \right\rangle}
\!\!\! {\cD}_{<i, i_2,..,j_L>} -(C+1)} \ ,
\end{equation}
and similarly for the restriction on the $j$ indices.

To proceed further, it is necessary to make an assumption about the
order parameters symmetry. The assumption made here is that of replica
symmetry in both the order parameters and the related conjugate
variables
\begin{eqnarray}
\label{eq:order_parameters_RS}
q_{\alpha, \beta.. \gamma} &=& a_{q} \int d x \ \pi(x) \
x^{l} \ , \ \widehat{q}_{\alpha, \beta.. \gamma}
= a_{\widehat{q}} \int d \hat{x} \ \widehat{\pi}(\hat{x}) \ 
\hat{x}^{l} \\ r_{\alpha, \beta.. \gamma} &=& a_{r}
\int d y \ \rho(y) \ y^{l}  \ , \ \widehat{r}_{\alpha,
\beta.. \gamma} = a_{\widehat{r}} \int d \hat{y} \
\widehat{\rho}(\hat{y}) \ \hat{y}^{l} \ , \nonumber
\end{eqnarray}
where $l$ is the number of replica indices, $a_{*}$ are normalisation
coefficients, and $\pi(x), \widehat{\pi}(\hat{x}), \rho(y)$ and
$\widehat{\rho}(\hat{y})$ represent probability distributions.
Unspecified integrals are over the range $[-1,+1]$. 
This ansatz is supported by the facts that (i) the current system 
is free of frustration and (ii) there has never been 
observed replica symmetry breaking at Nishimori's 
condition\cite{Nishimori_line}
which corresponds to using correct priors $F_s$ and $F_n$ 
in our case\cite{Iba}. 
The results obtained hereafter also support this ansatz. 
Extremizing the partition function with respect to distributions
$\pi(\cdot)$, $\hat{\pi}(\cdot)$, $\rho(\cdot)$ and $\hat{\rho}(\cdot)$, 
one then obtains the free energy per spin
\begin{eqnarray} 
    &\phantom{=}& f=-\frac{1}{N} \langle \ln
    {\cal Z} \rangle_{\xi,\zeta,\cD } \nonumber \\
    &=& \mbox{extr}_{\{ \pi, \hat{\pi}, \rho, \rho \} }
 \left \{ \frac{C}{K}\ln 2
    + C \int dx \ d \hat{x} \ \pi (x)  \ \hat{\pi}(\hat{x}) \ln 
    (1+x \hat{x})
    + \frac{CL}{K} \int dy \ d \hat{y} \ \rho(y) \
    \hat{\rho}(\hat{y}) \ln (1+y \hat{y}) \right . \nonumber \\
    &\phantom{=}&-\frac{C}{K}\int \left[ \prod_{k=1}^K
      dx_k \pi (x_k) \right]
    \left[ \prod_{l=1}^L dy_l \rho(y_l) \right]
    \ln \left[ 1+\prod_{k=1}^K x_k \prod_{l=1}^L y_l \right] \nonumber \\
    &\phantom{=}&- \int \left[ \prod_{k=1}^C d \hat{x}_k 
      \hat{\pi}(\hat{x}_k) \right]
    \left \langle \ln \left[
        e^{F_s \xi} \prod_{k=1}^C (1+\hat{x}_k)
        +e^{-F_s \xi}  \prod_{k=1}^C (1-\hat{x}_k)
      \right]
    \right \rangle _{\xi} \nonumber \\
&\phantom{=}& \left . - \frac{C}{K} \int \left[ \prod_{l=1}^C d \hat{y}_l 
      \hat{\rho}(\hat{y}_l) \right]
    \left \langle \ln \left[
        e^{F_n \zeta} \prod_{l=1}^L (1+\hat{y}_l)
        +e^{-F_n \zeta}  \prod_{l=1}^L (1-\hat{y}_l)
      \right]
    \right \rangle _{\zeta}
\right \},  
\label{eq:freeenergy} 
\end{eqnarray} 
where angled brackets with subscript $\bbox{\xi}$, $\bbox{\zeta}$ and 
$\cD$ denote averages over the message and noise distributions respectively, 
and sparse connectivity tensor $\cal D$. Message averages take the form 
\begin{equation}
\left \langle \cdots \right \rangle_\xi = 
\sum_{\xi=\pm 1} \frac{1+ \xi \tanh F_s}{2} \left ( \cdots \right )
\label{eq:average_xi}
\end{equation}
and similarly for $\left \langle \cdots \right \rangle_\zeta$. 
Details of the derivation are given in Appendix A.

Taking the functional variation of $f$ with respect to the
distributions $\pi$, $\hat{\pi}$, $\rho$ and $\hat{\rho}$, one obtains
the following saddle point equations
\begin{eqnarray}
    \pi(x)
    &=&
    \int \prod_{l=1}^{C-1}
    d \hat{x}_l \  \hat{\pi}(\hat{x}_l)
    \left\langle
      \delta
      \left(
        x-\tanh
        \left(
          \xi F_s+\sum_{l=1}^{C-1}
          \tanh^{-1}\hat{x}_l
        \right)
      \right)
    \right\rangle_{\xi}, \nonumber \\
    \hat{\pi}(\hat{x})
    &=&
    \int \prod_{l=1}^{K-1}
    d x_l \ \pi(x_l)
    \int \prod_{l=1}^L 
    d y_l \ \rho(y_l)
    \delta
    \left(
      \hat{x}-
        \prod_{l=1}^{K-1} x_l
        \prod_{l=1}^L y_l
    \right), \nonumber \\
    \rho(y)
    &=&
    \int \prod_{l=1}^{L-1}
    d \hat{y}_l \ \hat{\rho}(\hat{y}_l)
    \left\langle
      \delta
      \left(
        y-\tanh
        \left(
          \zeta F_n+\sum_{l=1}^{L-1}
          \tanh^{-1}\hat{y}_l
        \right)
      \right)
    \right\rangle_{\zeta}, \nonumber \\
    \hat{\rho}(\hat{y})
    &=&
    \int \prod_{l=1}^K
    d x_l \ \pi(x_l)
    \int \prod_{l=1}^{L-1}
    d y_l \ \rho(y_l)
    \delta
    \left(
      \hat{y}-
      \prod_{l=1}^K x_l
      \prod_{l=1}^{L-1} y_l
    \right). \label{eq:SPequations}
\end{eqnarray}
After solving these equations, the expectation of the overlap between 
the message $\mxi$ and the Bayesian optimal estimator 
(\ref{eq:bayesoptimal}), which serves as a performance measure, 
can be evaluated as 
\begin{eqnarray}
    m=\frac{1}{N} \left \langle \sum_{i=1}^N \xi_i \mbox{sign} 
\left \langle S_i \right \rangle_{\beta \to \infty} \right 
\rangle_{\mxi,\mzeta,\cD}=
\int dz \ \phi(z) \ \mbox{sign}(z), \label{eq:magnetization}
\end{eqnarray}
where 
\begin{eqnarray}
    \phi(z)=\int \left[ \prod_{l=1}^C d\hat{x}_l \ 
      \hat{\pi}(\hat{x}_l) \right] 
    \left\langle \delta 
      \left( 
        z-\tanh
        \left(
          F_s \xi+
          \sum_{i=1}^C
          \tanh^{-1}
          \hat{x}_i
        \right) 
      \right)
    \right\rangle_{\xi}. \label{eq:physical_field}
\end{eqnarray}
The derivation of Eqs.(\ref{eq:magnetization}) and (\ref{eq:physical_field}) 
is given in Appendix B. 

Examining the physical properties of the solutions for various
connectivity values exposes significant differences between the
various cases. In particular, these solutions fall into three
different categories: the cases of $K=1$ and general $L$ value, the
case of $K=L=2$ and all other parameter values where either $K\ge 3$
or $L\ge 3$ (and $K> 1$). We describe the results obtained for each
one of these cases separately.

\subsection{Analytical solution - the case of $K\ge 3$ or $L\ge 3$, $K>1$}
Results for the cases of $K\ge 3$ or $L\ge 3$, $K>1$ can be obtained
analytically and have a simple and transparent interpretation; we will
therefore focus first on this simple case.  For unbiased messages
(with $F_s=0$), one can easily verify that the ferromagnetic phase,
characterised by $m=1$, and the probability distributions
\begin{eqnarray}
        \pi(x)=\delta(x-1), \ \hat{\pi}(\hat{x})=\delta(\hat{x}-1), \
        \rho(y)=\delta(y-1), \ \hat{\rho}(\hat{y})=\delta(\hat{y}-1) \ ;
        \label{eq:ferro}
\end{eqnarray}
and the paramagnetic state of $m=0$ with the probability distributions
\begin{eqnarray}
        \pi(x)&=&\delta(x), \ \hat{\pi}(\hat{x})=\delta(\hat{x}),\
        \hat{\rho}(\hat{y})=\delta(\hat{y}), \nonumber \\
        \rho(y)&=&\frac{1+\tanh F_n}{2}\delta(y-\tanh F_n)
        +\frac{1-\tanh F_n}{2}\delta(y+\tanh F_n), \label{eq:para}
\end{eqnarray}
satisfy saddle point equations (\ref{eq:SPequations}). Other solutions
may be obtained numerically; here we have represented the
distributions by $10^3 - 10^4$ bins and iterated
Eqs.(\ref{eq:SPequations}) $100 - 500$ times with $10^5$ Monte Carlo
sampling steps for each iteration.  No solutions other than the above
two have been discovered.

The thermodynamically-dominant state is found
by evaluating the free energy of the two solutions using
Eq.(\ref{eq:freeenergy}), which yields 
\begin{eqnarray}
        f_{\mbox{ferro}}
        =
        -\frac{C}{K} F_n \tanh F_n = -\frac{1}{R} F_n \tanh F_n, 
        \label{eq:free_ferro}
\end{eqnarray}
for the ferromagnetic solution and 
\begin{eqnarray}
        f_{\mbox{para}}
        =
        \frac{C}{K}\ln 2 - \ln 2 - \frac{C}{K}\ln 2 \cosh F_n 
        = \frac{1}{R} \ln 2 - \ln 2 - \frac{1}{R} \ln 2 \cosh F_n, 
        \label{eq:free_para}
\end{eqnarray}
for the paramagnetic solution. 

Figure \ref{landscape_both}(a) describes schematically the nature of
the solutions for this case, in terms of the free energy and the
magnetisation obtained, for various flip rate probabilities. The
difference between the free energies of Eqs.(\ref{eq:free_ferro}) and
(\ref{eq:free_para})
\begin{eqnarray}
        f_{\mbox{ferro}}-f_{\mbox{para}}
        =\frac{\ln 2}{R} \left [ R-1+H_2(p) \right ], 
\label{eq:freeenegy_difference}
\end{eqnarray}
vanishes in the boundary between the two phase 
\begin{eqnarray}
        R_c=1-H_2(p)=1+p \log_2(p)+(1-p) \log_2(1-p), 
\label{eq:shannon_bound}
\end{eqnarray}
which coincides with Shannon's channel capacity. 

Equation (\ref{eq:shannon_bound}) indicates that all constructions
with either $K \ge 3$ or $L \ge 3$ (and $K>1$) can potentially realize
error-free data transmission for $R< R_c$ in the limit where both
message and codeword lengths $N$ and $M$ become infinite, thus
saturating Shannon's bound.

\subsection{The case of $K=L=2$}
All codes with either $K=3$ or $L=3$, $K>1$ potentially saturate
Shannon's bound and are characterised by a first order phase
transition between the ferromagnetic and paramagnetic solutions.  On
the other hand, numerical investigation based on Monte Carlo methods
indicates of significantly different physical characteristics for
$K=L=2$ codes shown in Fig.\ref{landscape_both}(b).

At the highest noise level, the paramagnetic solution (\ref{eq:para})
gives the unique extremum of the free energy until noise level reaches
the first critical point $p_1$, at which the ferromagnetic solution
(\ref{eq:ferro}) of higher free energy appears to be locally stable.
As the noise level decreases, a second critical point $p_2$ appears,
where the paramagnetic solution becomes unstable and a sub-optimal
ferromagnetic solution and its mirror image emerge.  Those solutions
have lower free energy than the ferromagnetic solution until the noise
level reaches the third critical point $p_3$.  Below $p_3$, the
ferromagnetic solution becomes the global minimum of the free
energy, while the sub-optimal ferromagnetic solutions still remain 
locally stable.  However, the sub-optimal solutions
disappear at the spinodal point $p_s$ and the ferromagnetic solution
(and its mirror image) becomes the unique stable solution of the saddle
point Eqs.(\ref{eq:SPequations}) as shown by the numerical
investigation for all $p <p_s$.

The analysis implies that $p_3$, the critical noise level below which
the ferromagnetic solution becomes thermodynamically dominant, is
lower than $p_c=H_2^{-1}(1-R)$ which corresponds to Shannon's bound.
Namely, $K=L=2$ does not saturate Shannon's bound in contrast to
$K\ge3$ codes even if optimally decoded.  Nevertheless, it turns out
that the free energy landscape, for noise levels $0< p < p_s$, offers
significant advantages in the decoding dynamic comparing to that of
other codes ($K\ge 3$ or $L\ge 3$, $K>1$).

\subsection{General $L$ codes with $K=1$}

The particular choice of $K=1$, independently of the value chosen for
$L$, exhibits a different behaviour presented schematically in
Fig.\ref{landscape_both}(c); also in this case there are no simple
analytical solutions and all solutions in this scenario, except for
the ferromagnetic solution, have been obtained numerically.  The first
important difference to be noted is that the paramagnetic state
(\ref{eq:para}) is no longer a solution of the saddle point equations
(\ref{eq:SPequations}) and is being replaced by a sub-optimal
ferromagnetic state. Convergence to the perfect solution of $m=1$ can
only be guaranteed for corruption rates smaller than that of the
spinodal point, marking the maximal noise level for which only the
ferromagnetic solution exists, $p<p_s$.

\begin{figure}
\vspace*{1.5cm}
\begin{center}
\epsfig{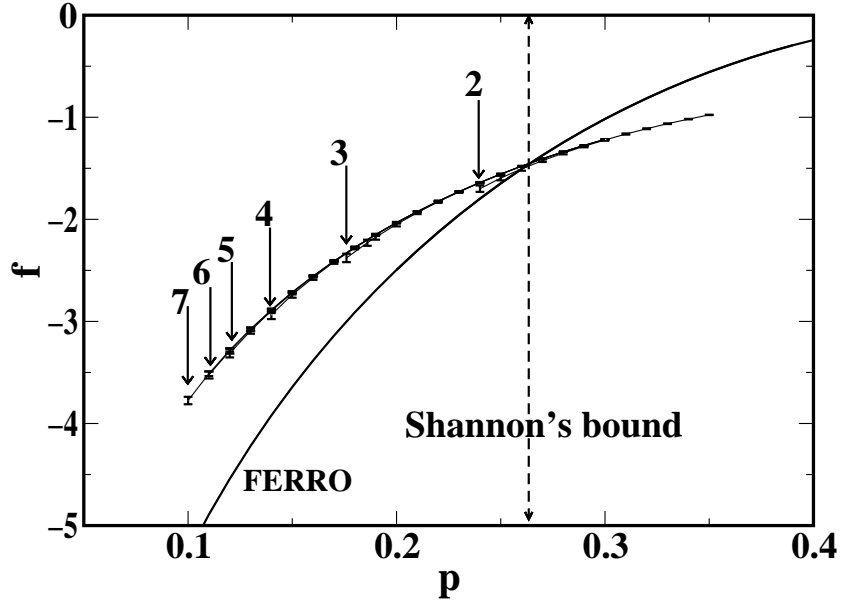}
\end{center}
\vspace*{2.5cm}
\caption{Free energies obtained by solving the analytical equations
 using Monte-Carlo integrations for $K=1$, $R=1/6$ and several values
 of $L$. Full lines represent the ferromagnetic free energy (FERRO,
 higher on the right) and the suboptimal ferromagnetic free energy
 (higher on the left) for values of $L=1,...,7$. The dashed line
 indicates Shannon's bound and the arrows represent the spinodal point
 values $p_s$ for $L=2,...,7$. The thermodynamic transition is very
 close, but bellow, the channel capacity ($p_1\approx 0.261$ against
 $p_c\approx 0.264$ at $R=1/6$).}
\label{k1lx}
\end{figure}

The $K=1$ codes do not saturate Shannon's bound in general; however,
we have found that at rates $R<1/3$ they outperform the $K=L=2$ code
while offering slightly improved dynamical (decoding)
properties. Studying the free energy in this case shows that as the
corruption rate increases, sub-optimal ferromagnetic solutions (stable
and unstable) emerge at the spinodal point $p_s$. When the noise
increases further this sub-optimal state becomes the global minimum at
$p_1$, dominating the system's thermodynamics.  The transition at
$p_1$ must occur at noise levels lower or equal to the value predicted
by Shannon's bound. In Fig.\ref{k1lx} we show free energy values
computed for a given code rate and several values of $L$, marking
Shannon's bound by a dashed line; it is clear that the thermodynamical
transition observed numerically (i.e. the point where the
ferromagnetic free energy equals the sub-optimal ferromagnetic free
energy) is bellow, but very close, to the channel capacity. It implies
that these codes also do not quite saturate Shannon's bound if
optimally decoded but get quite close to it.
 
\section{Decoding: Belief propagation/TAP approach}
The Bayesian message estimate (\ref{eq:bayesoptimal}) potentially
provides the optimal retrieval of the original messages.  However, it
is computationally difficult to follow the prescription exactly as it
requires a sum over ${\cal O}(2^N)$ terms.  Belief
propagation\cite{Frey,Pearl} (BP) can be used for obtaining an
approximate estimate.  It was recently shown\cite{us_TAP_sourlas} that
the BP algorithm can be derived, at least in the current context, from the
TAP approach\cite{TAP} to diluted systems in statistical mechanics.

Both algorithms (BP/TAP) are iterative methods which effectively
calculate the marginal posterior probabilities ${\cal P}(S_i |
\J,C_s,C_n)=\sum_{\{ \{S_{k \ne i}\}, \mtau \}} {\cal P}(\mS,\mtau |
\J,C_s,C_n)$ and ${\cal P}(\tau_j | \J,C_s,C_n)=\sum_{\{\mS, \{
\tau_{k \ne j} \} \}} {\cal P}(\mS, \mltau | \J,C_s,C_n)$ based on the
following three assumptions:
\begin{enumerate}
\item The posterior distribution is factorizable with respect to 
dynamical variables $S_{i=1,\ldots,N}$ and $\tau_{j=1,\ldots,M}$. 
\item The influence of check $J_{\mu=1,\ldots,M}$ 
on a specific site $S_i$ (or $\tau_j$) is also factorizable. 
\item The contribution of a single variables $S_{i=1,\ldots,N}$, 
$\tau_{j=1,\ldots,M}$ and $J_{\mu=1,\ldots,M}$ to the macroscopic 
variables is small and can be isolated. 
\end{enumerate}
Parameterising pseudo-marginal posteriors and 
marginalized conditional probabilities as
\begin{eqnarray}
{\cal P}(S_i | \{ J_{\nu \ne \mu} \}, C_s,C_n)&=&
 \frac{1+m_{\mu i }^S S_i}{2},  \quad 
{\cal P}(\tau_j | \{ J_{\nu \ne \mu} \}, C_s,C_n)=
\frac{1+m_{\mu j }^n \tau_j}{2}, \\
{\cal P}(J_\mu| S_i, \{J_{\nu \ne \mu}\}, C_s,C_n) &\sim& 
\frac{1+\hat{m}_{\mu i }^S S_i}{2}, \quad 
{\cal P}(J_\mu| \tau_j, \{J_{\nu \ne \mu}\}, C_s,C_n) \sim 
\frac{1+\hat{m}_{\mu j }^n \tau_j}{2}, 
\end{eqnarray}
the above assumptions provide a set of self-consistent equations
\cite{us_TAP_sourlas,Vicente}
\begin{eqnarray}
m^S_{\mu l} = \tanh\left (F_s+\sum_{\nu \in {\cal M}_S (l)/\mu} 
\tanh^{-1}(\hat{m}^S_{\nu l}) \right ), \quad 
m^n_{\mu l} = \tanh\left (F_n+\sum_{\nu \in {\cal M}_n (l)/\mu} 
\tanh^{-1}(\hat{m}^n_{\nu l}) \right ). \label{eq:hat_to_m}
\end{eqnarray}
and 
\begin{eqnarray}
\hat{m}^S_{\mu l} = J_\mu \prod_{k \in {\cal L}_S(\mu)/l} m^S_{\mu k}
\prod_{j \in {\cal L}_n(\mu)} m^n_{\mu j}, \quad
\hat{m}^n_{\mu l} = J_\mu \prod_{k \in {\cal L}_S(\mu)} m^S_{\mu k}
\prod_{j \in {\cal L}_n(\mu)/l} m^n_{\mu j}. \label{eq:m_to_hat}
\end{eqnarray}
Here, ${\cal M}_s(l)$ and ${\cal M}_n(l)$ indicate the set of all
indices of non-zero components in the $l$-th column of the sparse
matrices $C_s$ and $C_n$, respectively.  Similarly, ${\cal L}_s(\mu)$
and ${\cal L}_n(\mu)$ denote the set of all indices of non-zero
components in $\mu$-th row of the sparse matrices $C_s$ and $C_n$,
respectively.  The notation ${\cal L}_s(\mu)/l$ represents the set of
all indices belonging to ${\cal L}_s(\mu)$ except the index $l$.

Equations (\ref{eq:hat_to_m}) and (\ref{eq:m_to_hat}) are solved
iteratively using the appropriate initial conditions.  After obtaining
a solution to all $m_{\mu l}$ and $\hat{m}_{\mu l}$, an approximated
posterior mean can be calculated as
\begin{eqnarray}
m^S_{i}=\tanh \left ( F_s + 
\sum_{\mu \in {\cal M}_S (l)} \tanh^{-1} (\hat{m}^S_{\mu i}) \right ), 
\label{sec5:eqn3}
\end{eqnarray}
which provides an approximation to the Bayes-optimal estimator
(\ref{eq:bayesoptimal}) in the form of $\hat{\xi}^B=\mbox{sign} (m^S_{i})$.

Notice that the rather vague meaning of the fields distributions
introduced in the previous section becomes clear by introducing the
new variables $x=\xi_i m_{\mu i}^S$, $\hat{x}=\xi_i \hat{m}_{\mu
i}^S$, $y=\zeta_j m_{\mu j}^n$ and $\hat{y}=\zeta_j \hat{m}_{\mu
j}^n$\cite{Vicente}.  If one considers that these variables are
independently drawn from the distributions $\pi(x)$,
$\hat{\pi}(\hat{x})$, $\rho(y)$ and $\hat{\rho} (\hat{y})$, the
replica symmetric saddle point equations (\ref{eq:SPequations}) are
recovered from the BP/TAP equations (\ref{eq:hat_to_m}) and
(\ref{eq:m_to_hat}).  This connection can be extended to the free
energy as equations (\ref{eq:hat_to_m}) and (\ref{eq:m_to_hat})
extremize the TAP free energy
\begin{eqnarray}
    f_{\mbox{TAP}}(\{ \bm \}, \{\hat{\bm } \})
    &=&\frac{M}{N}\ln 2 +
    \frac{1}{N}\sum_{\mu=1}^M \sum_{i \in {\cal L}_S (\mu)} \ln
    \left(
      1+m_{\mu i}^S \hat{m}_{\mu i}^S
    \right)
    +\frac{1}{N}\sum_{\mu =1}^M \sum_{j \in {\cal L}_n (\mu)} \ln
    \left(
      1+m_{\mu j}^{n} \hat{m}_{\mu j}^{n} \right) \nonumber \\
    &\phantom{=}& - \frac{1}{N}\sum_{\mu=1}^M
    \ln \left(
      1+J_{\mu}\prod_{i \in {\cal L}_S (\mu)}m_{\mu i}^S
        \prod_{j \in {\cal L}_n (\mu)} m_{\mu j}^{n}\right) \nonumber \\
    &\phantom{=}&- 
    \frac{1}{N}\sum_{i=1}^N \ln
    \left[
      e^{F_s} \prod_{\mu \in {\cal M}_S (i)}
      \left(
        1+\hat{m}_{\mu i}^S
      \right)
      +e^{-F_s} \prod_{\mu \in {\cal M}_S (i)}
      \left(
        1-\hat{m}_{\mu i}^S
      \right)
    \right] \nonumber \\
    &\phantom{=}&-
    \frac{1}{N}\sum_{j=1}^M \ln
    \left[
      e^{F_n} \prod_{\mu \in {\cal M}_n (j)}
      \left(
        1+\hat{m}_{\mu j}^{n}
      \right)
      +e^{-F_n} \prod_{\mu \in {\cal M}_n (j)}
      \left(
        1-\hat{m}_{\mu j}^{n}
      \right)
    \right] \  .
\label{eq:TAP_freeenergy}
\end{eqnarray}
This expression may be used for selecting the thermodynamically
dominant state when Eqs.(\ref{eq:hat_to_m}) and (\ref{eq:m_to_hat})
have several solutions.

We have investigated the performance of the various codes using BP/TAP
equations as the decoding algorithm.  Solutions have been obtained by
iterating the equations (\ref{eq:hat_to_m}) and (\ref{eq:m_to_hat})
$100-500$ times under various initial conditions.  Since the system is
not frustrated, the dynamics converges within $10-30$ updates in most
cases except close to criticality. The numerical results mirror the 
behaviour predicted by the analytical solutions.

For either $K \ge 3$ or $L \ge 3$ ,$K>1$ codes, the ferromagnetic
solution
\begin{eqnarray}
m_{\mu i}^S=\xi_i, \quad \hat{m}_{\mu i}^S=\xi_i, \quad 
m_{\mu j}^n=\zeta_j, \quad \hat{m}_{\mu j}^n=\zeta_j, 
\label{eq:ferro_TAP}
\end{eqnarray}
which provides perfect decoding ($m=1$) and the paramagnetic solution
($m=0$)
\begin{eqnarray}
m_{\mu i}^S=0, \quad \hat{m}_{\mu i}^S=0, \quad 
m_{\mu j}^n=\tanh F_n =1-2p, \quad \hat{m}_{\mu j}^n=0 \ , 
\label{eq:para_TAP}
\end{eqnarray}
are obtained in various runs depending on the initial conditions (the
message is assumed unbiased resulting in $F_s=0$).  However, it is
difficult to set the initial conditions within the basin of attraction
of the ferromagnetic solution without prior knowledge about the
transmitted message $\mxi$.

\begin{figure}
\vspace*{1.5cm}
\begin{center}
\epsfig{file=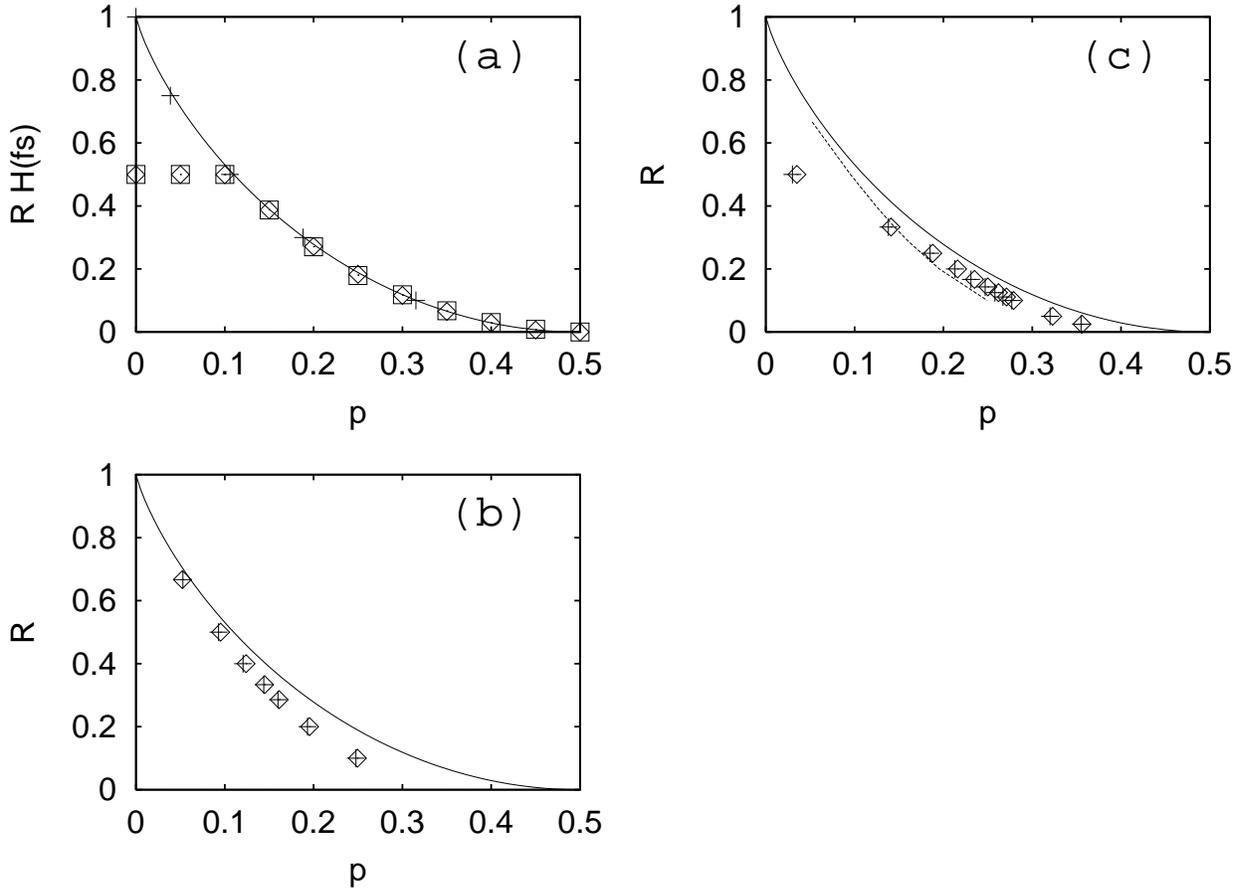,angle=0,width=170mm}
\end{center}
\caption{Critical code rate as a function of the flip rate $p$,
obtained from numerical solutions and the TAP approach ($N\!=\!
10^4$), and averaged over 10 different initial conditions with error
bars much smaller than the symbols size.
(a) Numerical solutions for $K\!  =\! L \!=\!3$, $C\!=\! 6$ and
varying input bias $f_{s}$ ($\Box$) and TAP solutions for both
unbiased ($+$) and biased ($\Diamond$) messages; initial conditions
were chosen close to the analytical ones. The critical rate is
multiplied by the source information content to obtain the maximal
information transmission rate, which clearly does not go beyond
$R\!=\!3/6$ in the case of biased messages; for unbiased patterns
$H_{2}(f_{s})\!=\!1$.
(b) For the unbiased case of $K\!  =\! L \!=\!2$; initial conditions
for the TAP ($+$) and the numerical solutions ($\Diamond$) were chosen
to be of almost zero magnetisation.
(c) For the case of $K=1$, $L=2$ and unbiased messages. We show
numerical solutions of the analytical equations ($\Diamond$) and those
obtained by the TAP approach ($+$).  The dashed line indicates the
performance of $K=L=2$ codes for comparison.  Codes with $K=1$, $L=2$
outperform $K=L=2$ for code rates $R<1/3$.}
\label{diagram}
\end{figure}

Biased coding is sometimes used for alleviating this difficulty
\cite{MN}.  Using a redundant source of information (equivalent to the
introduction of a non-zero field $F_s$ in the statistical physics
description), one effectively increases the probability of the initial
conditions being closer to the ferromagnetic solution.  The main
drawback of this method is that the information per transmitted bit is
significantly reduced due to this redundancy.  In order to investigate
how the maximum performance is affected by transmitting biased
messages, we have evaluated the critical information rate (i.e., code
rate $\times H_2(f_s=(1+\tanh F_s)/2)$, the source redundancy), below
which the ferromagnetic solution becomes thermodynamically dominant
[Fig.\ref{diagram}(a)].  The data were obtained by the BP/TAP method
(diamonds) and numerical solutions of from replica framework (square);
the dominant solution in the BP/TAP results, was selected by using the
free energy (\ref{eq:TAP_freeenergy}).  Numerical solutions have been
obtained using $10^3-10^4$ bin models for each distribution and had
been run for $10^5$ steps per noise level.  The various results are
highly consistent and practically saturate Shannon's bound for the
same noise level. However, it is important to point out that close to
Shannon's limit, prior knowledge on the original message is required
for setting up appropriate initial conditions that ensure convergence
to the ferromagnetic solution; such prior knowledge is not available
in practice.

Although $K,L \ge 3$ codes seem to offer optimal performance when
highly biased messages are transmitted, this seems to be of little
relevance in most cases, characterised by the transmission of
compressed unbiased messages or only slightly biased messages. 
 In this sense, $K=L=2$ and $K=1$ codes
can be considered more practical as the BP/TAP dynamics of these codes
exhibit unique convergence to the ferromagnetic solution (or mirror
image in the $K=L=2$ case) from {\em any} initial condition up to a
certain noise level.  This property results from the fact that the
corresponding free energies have no local minima other than the
ferromagnetic solution below $p_s$.

In figures \ref{diagram}(b) and (c) we show the value of $p_s$ for
the cases of $K=L=2$ and $K=1$, $L=2$ respectively, evaluated by
numerical solutions from the replica framework (diamonds) and 
 using the BP/TAP method.

The case of $K=L=2$ shows consistent successful decoding for the code
rates examined and up to noise levels slightly below, but close to,
Shannon's bound.  It should be emphasised here that initial conditions
are chosen almost randomly in the BP/TAP method, with a very slight
bias of ${\cal O}(10^{-12})$ in the initial magnetisation.  This
result suggests using $K=L=2$ codes (or similar), rather than $K,L \ge
3$ codes, although the latter may potentially have better equilibrium
properties.

In Fig.\ref{diagram}(c) we show that for code rates $R<1/3$, codes
parametrised by $K=1$ and $L=2$ outperform $K=L=2$ codes with one
additional advantage: Due to the absence of mirror symmetries these
codes converge to the ferromagnetic state much faster, and there is no
risk of convergence to the mirror solution. The difference in
performance becomes even larger as the code rate decreases. Higher
code rates will result in performance deterioration due to the low
connectivity, eventually bringing the system below the percolation
threshold.

\begin{figure}
\vspace*{1.5cm}
\begin{center} 
\epsfig{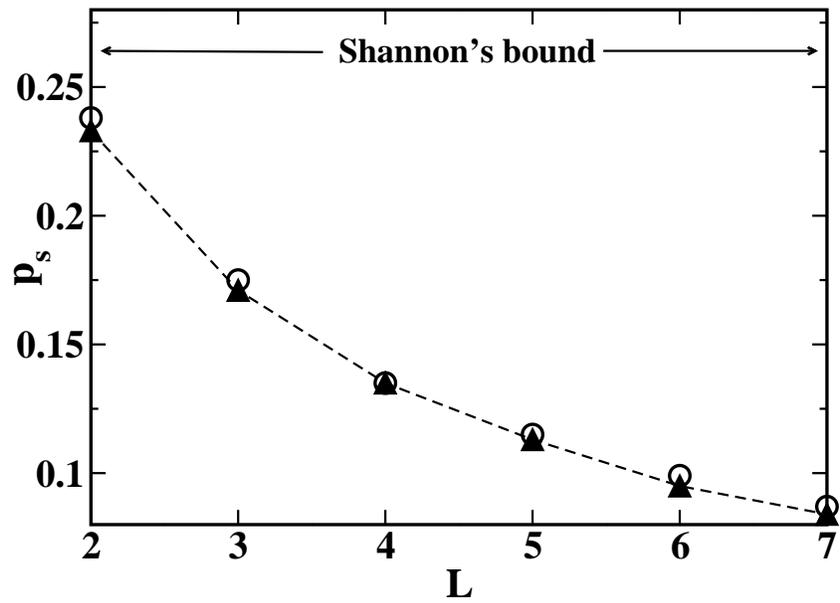}
\end{center}
\vspace*{2.5cm}
\caption{The spinodal point noise level $p_s$ for $K=1$, $R=1/6$ and several
choices of $L$.  Numerical solutions are denoted by circles and TAP
decoding solutions ($N\!=\!10^4$) by black triangles.  }
\label{k1lxperformance}
\end{figure}

In Fig.\ref{k1lxperformance} we examine the dependence of the noise
level of the spinodal point $p_s$ on the value of $L$, and show that
the choice of $L=2$ is optimal within this family.  Codes with $L=1$
have very poor error-correction capabilities as their Hamiltonian
(\ref{eq:hamiltonian}) corresponds to the Mattis model, which is
equivalent to a simple ferromagnet in a random field attaining
magnetisation $m=1$ only in the noiseless case.

\section{Reducing  Encoding Costs}
The BP/TAP algorithm already offers an efficient decoding method,
which requires ${\cal O}(N)$ operations; however, the current encoding
scheme includes three costly processes: (a) The computational cost of
constructing the generating matrix $C_n^{-1} C_s$ requires ${\cal
O}(N^3)$ operations for inverting the matrix $C_n$ and ${\cal O}(N^2)$
operations for the matrix multiplication.  (b) The memory allocation
for generating the matrix $C_n^{-1} C_s$ scales as ${\cal O}(N^2)$
since this matrix is typically dense. (c) The encoding itself
$\mz_0=C_n^{-1}C_s \mxi$ (mod $2$) requires ${\cal O}(N^2)$
operations.

These computational costs become significant when long messages
$N=10^4 \sim 10^5$ are transmitted, which is typically the case for
which Gallager-type codes are being used. This may require long
encoding times and may delay the transmission.

These problems may be solved by utilising systematically constructed
matrices instead of random ones, of some similarity to the
constructions of \cite{Kanter_Saad}. Here, we present a simple method
to reduce the computational and memory costs to ${\cal O}(N)$ for
$K=L=2$ and $K=1$, $L=2$ codes.  Our proposal is mainly based on using
a specific matrix for $C_n$,
\begin{eqnarray}
\bar{C}_n=\left (
\begin{array}{cccccccc}
1 & 0 & 0 & 0 & \cdots & 0 & 0 \cr
1 & 1 & 0 & 0 & \cdots & 0 & 0 \cr
0 & 1 & 1 & 0 & \cdots & 0 & 0 \cr
0 & 0 & 1 & 1 & \cdots & 0 & 0 \cr
\vdots & \vdots & \vdots & \ddots & \ddots & \vdots & \vdots \cr
0 & 0 & 0 & 0 & \cdots & 1 & 1
\end{array}
\right ) \ , 
\label{eq:regular_Cn}
\end{eqnarray} 
instead of a randomly-constructed one.  For $C_s$, we use a random
matrix of $K=2$ (or $K=1$) non-zero elements per row as before.

The inverse (mod 2) of $\bar{C}_n^{-1}$ becomes the lower 
triangular matrix
\begin{eqnarray}
\bar{C}_n^{-1}=\left (
\begin{array}{cccccccc}
1 & 0 & 0 & 0 & \cdots & 0 & 0 \cr
1 & 1 & 0 & 0 & \cdots & 0 & 0 \cr
1 & 1 & 1 & 0 & \cdots & 0 & 0 \cr
1 & 1 & 1 & 1 & \cdots & 0 & 0 \cr
\vdots & \vdots & \vdots & \ddots & \ddots & \vdots & \vdots \cr
1 & 1 & 1 & 1 & \cdots & 1 & 1
\end{array}
\right ) \ . 
\label{eq:regular_Cn_inv}
\end{eqnarray} 
This suggests that encoding the message $\mxi$ into 
a codeword $\mz^0$ would require  only   
${\cal O}(N)$ operations by carrying it out in two steps 
\begin{eqnarray}
t_\mu &=& (C_s \mxi)_\mu \quad \mbox{(mod $2$)}, 
\quad \mbox{for $\mu=1,2,\ldots,M$}, 
\label{eq:fast1} \\
z^0_\mu &=& (\bar{C}_n^{-1} \mt)_\mu= 
z^0_{\mu-1}+t_\mu \quad \mbox{(mod $2$)}, \quad \mbox{for $\mu=2,\ldots,M$}, 
\label{eq:fast2}
\end{eqnarray}
with $z^0_1=t_1$.  Both steps require ${\cal O}(N)$ operations due to
the sparse nature of $C_s$.  In addition, the required memory
resources are also reduced to ${\cal O} (N)$ since only the sparse
matrix $C_s$ should be stored.

The possible drawback of using the systematic matrix
(\ref{eq:regular_Cn}) is a deterioration in the error correction
ability.  We have examined numerically the performance of new
construction to discover, to our surprise, that it is very similar to
that of random matrix based codes as shown in Table \ref{performance}.
Although our examination is only limited to BSC and i.i.d. messages,
it seems to suggest that some deterministically constructed matrices
may be implemented successfully in practice.

\section{Summary}
In this paper, we have investigated the typical performance of the MN
codes, a variation of Gallager-type error-correcting codes, by mapping
them onto Ising spin models and making use of the established methods
of statistical physics.  We have discovered that for a certain choice
of parameters, either $K \ge 3$ or $L \ge 3$, $K>1$ these codes
potentially saturate the channel capacity, although this cannot be
used efficiently in practice due to the decrease in the basin of
attraction which typically diverts the decoding dynamics towards the
undesired paramagnetic solution. 

Codes with $K=2$ and $L=2$ show close to optimal performance while
keeping a large basin of attraction, resulting in more practical
codes.  Constructions of the form $K=1$, $L=2$ outperform the $K=L=2$
codes for code rates $R<1/3$, having improved dynamical
properties.

These results are complementary to those obtained so far by the
information theory community and seem to indicate that worst-case
analysis can be, in some situations, too pessimistic when compared to
the typical performance results.

Beyond the theoretical aspects, we proposed an efficient method for
reducing the computational costs and the required memory allocation by
using a specific construction of the matrix $C_n$. These codes are
highly attractive and provide lower computational costs for both
encoding and decoding.

Various aspects that remain to be studied include a proper analysis of
the finite size effects for rates below and above the channel
capacity, which are of great practical relevance; and the use of
statistical physics methods for optimising the matrix constructions.

\section*{Acknowledgement}
Support by the
JSPS RFTF program (YK), The Royal Society and EPSRC grant GR/N00562 (DS) is
acknowledged.

\newpage

\appendix
\section{Replica Free Energy}
The purpose of this appendix is to derive the averaged free energy per 
spin (\ref{eq:freeenergy}). 
Applying the gauge transformation
\begin{eqnarray}
J_{\mu} &\to& J_{\mu} 
        \prod_{i \in {\cal L}_s(\mu)}
        \xi_i
        \prod_{j \in {\cal L}_n(\mu)}
        \zeta_j
        =1 \nonumber \\
S_i &\to& S_i \xi_i \nonumber \\
\tau_j &\to& \tau_j \zeta_j,
\label{eq:gauge_appendix}
\end{eqnarray}
to eq. (\ref{eq:partition_function}), one may rewrite the partition
function in the form
\begin{eqnarray}
    &&{\cal Z}(\bbox{\xi},\bbox{\zeta},{\cal D})
    =
    \sum_{\bbox{S},\bbox{\tau}}
    \exp
    \left(
      F_s \sum_{i=1}^N \xi_i S_i
      +
      F_n \sum_{j=1}^M \zeta_j \tau_j
    \right) \nonumber \\
    &\times& \prod_{\langle i_1,\cdots ,i_K;j_1,\cdots,j_L \rangle}
    \left[
      1-{\cal D}_{\langle i_1,\cdots ,i_K;j_1,\cdots,j_L \rangle}
      +{\cal D}_{\langle i_1,\cdots ,i_K;j_1,\cdots,j_L \rangle}
      \frac{1}{2}
      \left(
        1+
        S_{i_1}\cdots S_{i_K}
        \tau_{j_1}\cdots \tau_{j_L}
      \right)
    \right]. \label{eq:gauged_pfunction}
\end{eqnarray}
Using the replica method, one calculates the quenched average of the
$n$-th power of the partition function given by
\begin{eqnarray}
    &&\langle 
    {\cal Z}(\bbox{\xi},\bbox{\zeta},{\cal D})^n 
    \rangle_{\bbox{\xi},\bbox{\zeta},{\cal D}}
    =
    \sum_{\bbox{S}^1 \cdots \bbox{S}^n}
    \sum_{\bbox{\tau}^1 \cdots \bbox{\tau}^n}
    \left\langle
      \exp
      \left(
        F_s \sum_{i=1}^N \xi_i \sum_{\alpha=1}^n S_i^{\alpha}
      \right)
    \right\rangle_{\bbox{\xi}}
    \left\langle
      \exp
      \left(
        F_n \sum_{j=1}^M \zeta_j \sum_{\alpha=1}^n \tau_i^{\alpha}
      \right)
    \right\rangle_{\bbox{\zeta}} \nonumber \\
    &\times&
    \left\langle
      \prod_{\langle i_1,\cdots ,i_K;j_1,\cdots,j_L \rangle}
      \prod_{\alpha=1}^n
      \left\{
        1+\frac{1}{2}
        {\cal D}_{\langle i_1,\cdots ,i_K;j_1,\cdots,j_L \rangle}
        \left(
          S_{i_1}^{\alpha} \cdots S_{i_K}^{\alpha}
          \tau_{j_1}^{\alpha} \cdots \tau_{j_L}^{\alpha}
          -1
        \right)
      \right\}
    \right\rangle_{\cal D}, \label{eq:nth_partition_function}
\end{eqnarray}
where averages with respect to $\mxi$ can be  easily performed 
\begin{eqnarray}
    \left\langle
      \exp
      \left(
        F_s \sum_{i=1}^N \xi_i \sum_{\alpha=1}^n S_i^{\alpha}
      \right)
    \right\rangle_{\bbox{\xi}}
    &=&
    \prod_{i=1}^N
    \left[
      \left (\frac{1+\tanh F_s}{2} \right ) 
        e^{F_s \sum_{\alpha=1}^n S_i^{\alpha}}
      +
      \left (\frac{1-\tanh F_s}{2} \right ) 
        e^{
        -F_s \sum_{\alpha=1}^n S_i^{\alpha}}
   \right] \cr
&=& \prod_{i=1}^N 
\left \langle \exp \left (\xi F_s \sum_{a=1}^n S_i^a \right ) \right 
\rangle_\xi, 
\label{eq:xi_average}
\end{eqnarray}
and similarly for $\langle \cdots \rangle_{\bbox{\zeta}}$.
The main problem is in averages over the sparse tensor realisations $\cal D$, 
which have complicated constraints. Following the procedure 
introduced by Wong and Sherrington\cite{Wong}, it is being rewritten as 
\begin{eqnarray}
    &&
    \left\langle
      \prod_{\langle i_1,\cdots ,i_K;j_1,\cdots,j_L \rangle}
      \prod_{\alpha=1}^n
      \left[
        1+\frac{1}{2}
        {\cal D}_{\langle i_1,\cdots ,i_K;j_1,\cdots,j_L \rangle}
        \left(
          S_{i_1}^{\alpha} \cdots S_{i_K}^{\alpha}
          \tau_{j_1}^{\alpha} \cdots \tau_{j_L}^{\alpha}
          -1
        \right)
      \right]
    \right\rangle_{\cal D} \nonumber \\
    &=&
    {\cal N}^{-1} \sum_{\cal D} 
    \prod_{i=1}^N
    \delta
    \left(
      \sum_{\langle i_1,\cdots ,i_K;j_1,\cdots,j_L \rangle}
      {\cal D}_{\langle i,i_2,\cdots ,i_K;j_1,\cdots,j_L \rangle}
      -C
    \right)
    \prod_{j=1}^M
    \delta
    \left(
      \sum_{\langle i_1,\cdots ,i_K;j_1,\cdots,j_L \rangle}
      {\cal D}_{\langle i_1,\cdots ,i_K;j,j_2,\cdots,j_L \rangle}
      -L
    \right) \nonumber \\
    &\phantom&\times
    \prod_{\langle i_1,\cdots ,i_K;j_1,\cdots,j_L \rangle}
    \prod_{\alpha=1}^n
    \left[
      1+\frac{1}{2}
      {\cal D}_{\langle i_1,\cdots ,i_K;j_1,\cdots,j_L \rangle}
      \left(
        S_{i_1}^{\alpha} \cdots S_{i_K}^{\alpha}
         \tau_{j_1}^{\alpha} \cdots \tau_{j_L}^{\alpha}
        -1
      \right)
    \right], 
\label{eq:wong}
\end{eqnarray}
where $\delta(\cdots)$ represents Dirac's $\delta$-function and 
\begin{eqnarray}
    {\cal N}
    =
    \sum_{\cal D} \prod_{i=1}^N
    \delta
    \left(
      \sum_{\langle i_2,\cdots,i_K;j_1,\cdots,j_L \rangle}
      {\cal D}_{\langle i,i_2,\cdots,i_K;j_1,\cdots,j_L \rangle}-C
    \right)
    \prod_{j=1}^M
    \delta
    \left(
      \sum_{\langle i_1,\cdots,i_K;j_2,\cdots,j_L \rangle}
      {\cal D}_{\langle i_1,\cdots,i_K;j,j_2,\cdots,j_L \rangle}-L
    \right) \label{eq:normalization}
\end{eqnarray}
represents the normalisation constant. 

We first evaluate this normalisation constant using the integral
representation of the $\delta$-function and
Eq.(\ref{eq:normalization}), to obtain
\begin{eqnarray}
    {\cal N}
    &=&
    \sum_{\cal D} \prod_{i=1}^N
    \delta
    \left(
      \sum_{\langle i_2,\cdots,i_K;j_1,\cdots,j_L \rangle}
      {\cal D}_{\langle i,i_2,\cdots,i_K;j_1,\cdots,j_L \rangle}-C
    \right)
    \prod_{j=1}^M
    \delta
    \left(
      \sum_{\langle i_1,\cdots,i_K;j_2,\cdots,j_L \rangle}
      {\cal D}_{\langle i_1,\cdots,i_K;j,j_2,\cdots,j_L \rangle}-L
    \right)\nonumber \\
    &=&
    \sum_{\cal D} \prod_{i=1}^N
    \left\{
      \int_0^{2\pi} \frac{d \lambda_i}{2 \pi}
      \exp
      \left[
        i \lambda_i
        \left(
          \sum_{\langle i_2,\cdots,i_K;j_1,\cdots,j_L \rangle}
          {\cal D}_{\langle i,i_2,\cdots,i_K;j_1,\cdots,j_L \rangle}
          -C
        \right)
      \right]
    \right\} \nonumber \\
    &\phantom&\times
    \prod_{j=1}^M
    \left\{
      \int_0^{2\pi} \frac{d \lambda_j}{2 \pi}
      \exp
      \left[
        i \lambda_j
        \left(
          \sum_{\langle i_1,\cdots,i_K;j_2,\cdots,j_L \rangle}
          {\cal D}_{\langle i_1,\cdots,i_K;j,j_2,\cdots,j_L \rangle}
          -L
        \right)
      \right]
    \right\} \nonumber \\
    &=&
    \prod_{i=1}^N
    \left\{
      \int_0^{2\pi} \frac{d \lambda_i}{2\pi}
      e^{-i C \lambda_i}
    \right\}
    \prod_{j=1}^M
    \left\{
      \int_0^{2\pi} \frac{d \nu_j}{2\pi}
      e^{-i L \nu_j}
    \right\} \nonumber \\
    &\phantom{=}&\times
    \sum_{\cal D}
    \prod_{i=1}^N
    \left\{
      \prod_{\langle i_2,\cdots,i_K;j_1,\cdots,j_L \rangle}
      e^{i \lambda_i {\cal D}_{\langle i,i_2,\cdots,i_K;j_1,\cdots,j_L 
    \rangle}}
    \right\}
    \prod_{j=1}^M
    \left\{
      \prod_{\langle i_1,\cdots,i_K;j_2,\cdots,j_L \rangle}
      e^{i \nu_j {\cal D}_{\langle i_1,\cdots,i_K;j,j_2,\cdots,j_L 
      \rangle}}
    \right\} \nonumber \\
    &=&
    \prod_{i=1}^N
    \left\{
      \int_0^{2\pi} \frac{d \lambda_i}{2\pi}
      e^{-i C \lambda_i}
    \right\}
    \prod_{j=1}^M
    \left\{
      \int_0^{2\pi} \frac{d \nu_j}{2\pi}
      e^{-i L \nu_j}
    \right\} \nonumber \\
    &\phantom{=}&\times
    \sum_{\cal D}
    \prod_{\langle i_1,\cdots,i_K;j_1,\cdots,j_L \rangle}
    \left\{
      \left(
        e^{i\lambda_{i_1}}\cdots e^{i \lambda_{i_K}}
        e^{i\nu_{j_1}}\cdots e^{i \nu_{j_L}}
      \right)^{{\cal D}_{\langle i_1,\cdots,i_K;j_1,\cdots,j_L \rangle}}
    \right\} \cr
&=& \prod_{i=1}^N
    \left\{
      \oint \frac{d Z_i}{2\pi i}Z_i^{-(C+1)}
    \right\}
    \prod_{j=1}^M
    \left\{
      \oint \frac{d Y_j}{2\pi i}Y_j^{-(L+1)}
    \right\}
    \prod_{\langle i_1,\cdots,i_K;j_1,\cdots,j_L \rangle}
    \left(
      1+
      Z_{i_1}\cdots Z_{i_K}
      Y_{j_1}\cdots Y_{j_L}
    \right), 
\label{eq:normalization_integral}
\end{eqnarray}
where we made use of the transformations $Z_i=e^{i \lambda_i},
Y_j=e^{i \nu_j}$, and carried out summations with respect to the
realisation of ${\cal D}$. Expanding the product on the right hand
side one obtains
\begin{eqnarray}
    &&\prod_{\langle i_1,\cdots ,i_K;j_1,\cdots,j_L \rangle}
    \left[
      1+
      \left(
        Z_{i_1}\cdots Z_{i_K}
        Y_{j_1}\cdots Y_{j_L}
      \right)
    \right] \nonumber \\
    &=&\exp
    \left[
      \sum_{\langle i_1,\cdots ,i_K;j_1,\cdots,j_L \rangle}
      \ln
      \left \{
        1+
        \left(
          Z_{i_1}\cdots Z_{i_K}
          Y_{j_1}\cdots Y_{j_L}
        \right)
      \right \}
    \right] \nonumber \\
&\simeq&\exp
    \left[
      \sum_{\langle i_1,\cdots ,i_K;j_1,\cdots,j_L \rangle}
      \left(
        Z_{i_1}\cdots Z_{i_K}
        Y_{j_1}\cdots Y_{j_L}
      \right)
    \right] \nonumber \\
&\simeq&     \exp
    \left[
      \frac{1}{K!}
      \left(
        \sum_{i=1}^N
        Z_i
      \right)^K
      \frac{1}{L!}
      \left(
        \sum_{j=1}^M
        Y_j
      \right)^L
    \right], \label{app1:eqn12}
\end{eqnarray}
in the thermodynamic limit. 
Using the identities
\begin{eqnarray}
    1=\int dq\;
    \delta
    \left(
      \sum_{i=1}^N
      Z_i-q
    \right),\quad
    1=\int dr \;
    \delta
    \left(
      \sum_{j=1}^M
      Y_j-r
    \right) \label{eq:delta_conjugate}
\end{eqnarray}
Eq. (\ref{eq:normalization_integral}) becomes
\begin{eqnarray}
    {\cal N}
    &=&
    \int dq \;\delta
    \left(
      \sum_{i=1}^N Z_i - q
    \right)
    \int dr \; \delta
    \left(
      \sum_{j=1}^M Y_j -r
    \right) \nonumber \\
    &\phantom{=}&\times
    \prod_{i=1}^N
    \left\{
      \oint \frac{d Z_i}{2\pi i}Z_i^{-(C+1)}
    \right\}
    \prod_{j=1}^M
    \left\{
      \oint \frac{d Y_j}{2\pi i}Y_j^{-(L+1)}
    \right\}
    \exp
    \left(
      \frac{q^K}{K!}
      \frac{r^L}{L!}
    \right) \nonumber \\
    &=&
    \int dq
    \int \frac{d \hat{q}}{2 \pi i}
    \exp
    \left[
      \hat{q}
      \left(
        \sum_{i=1}^N Z_i - q
      \right)
    \right]
    \int dr
    \int \frac{d \hat{r}}{2 \pi i}
    \exp
    \left[
      \hat{r}
      \left(
        \sum_{j=1}^M Y_j - r
      \right)
    \right] \nonumber \\
    &\phantom{=}& \times
    \prod_{i=1}^N
    \left\{
      \oint \frac{d Z_i}{2\pi i}Z_i^{-(C+1)}
    \right\}
    \prod_{j=1}^M
    \left\{
      \oint \frac{d Y_j}{2\pi i}Y_j^{-(L+1)}
    \right\}
    \exp
    \left(
      \frac{q^K}{K!}
      \frac{r^L}{L!}
    \right) \nonumber \\
    &=&
    \int dq \int \frac{d \hat{q}}{2 \pi i}
    \int dr \int \frac{d \hat{r}}{2 \pi i}
    \exp
    \left(
      \frac{q^K}{K!}\frac{r^L}{L!}-q \hat{q}-r \hat{r}
    \right) \nonumber \\
    &\phantom{=}&\times
    \prod_{i=1}^N
    \left[
      \oint \frac{dZ_i}{2 \pi i} Z_i^{-(C+1)}
      \exp
      \left(
        \hat{q}Z_i
      \right)
    \right]
    \prod_{j=1}^M
    \left[
      \oint \frac{dY_j}{2 \pi i} Y_j^{-(L+1)}
      \exp
      \left(
        \hat{r}Y_j
      \right)
    \right] \ .  \label{app1:eqn18}
\end{eqnarray}
The contour integrals provide the following constants
\begin{eqnarray}
    \prod_{i=1}^N
    \left[
      \oint \frac{dZ_i}{2 \pi i} Z_i^{-(C+1)}
      \exp
      \left(
        \hat{q}Z_i
      \right)
    \right]
    =
    \left(
      \frac{\hat{q}^C}{C!}
    \right)^N,\quad
    \prod_{j=1}^M
    \left[
      \oint \frac{dY_j}{2 \pi i} Y_j^{-(L+1)}
      \exp
      \left(
        \hat{r}Y_j
      \right)
    \right]
    =
    \left(
      \frac{\hat{r}^L}{L!}
    \right)^M, \label{eq:contour_integrals}
\end{eqnarray}
respectively.  Applying the 
saddle point method to the remaining integrals, 
one obtains
\begin{eqnarray}
    {\cal N}
    =
    {\mbox{extr}}_{\{ q,\hat{q},r,\hat{r} \}}
    \left\{
      \exp
      \left[
        \frac{q^K}{K!}\frac{r^L}{L!}-q \hat{q} -r \hat{r}
        +NC \ln \hat{q} -N \ln (C!)
        +ML \ln \hat{r} -M \ln (L!)
      \right]
    \right\} \ , \label{eq:normalization_extremization}
\end{eqnarray}
which yields the following saddle point equations with respect to 
$q$, $r$, $\hat{q}$ and $\hat{r}$ 
\begin{eqnarray}
    q&=&\frac{NC}{\hat{q}},\quad
    r=\frac{ML}{\hat{r}} \nonumber \\
    \hat{q}
    &=&
    \frac{q^{K-1}}{(K-1)!}
    \frac{r^L}{L!},\quad 
    \hat{r}
    =
    \frac{r^{L-1}}{(L-1)!}
    \frac{q^K}{K!} \ , \label{eq:normalization_sp_equation}
\end{eqnarray}
providing the normalisation constant 
\begin{eqnarray}
    {\cal N}
    =
    \left(
      \frac{\hat{q}^C}{C!}
    \right)^N
    \left(
      \frac{\hat{r}^L}{L!}
    \right)^M
    \exp
    \left(
      \frac{q^K}{K!}\frac{r^L}{L!}-q \hat{q} -r \hat{r}
    \right) \ . \label{eq:extremized_normalization}
\end{eqnarray}

Equation (\ref{eq:wong}) can be evaluated similarly. Following a similar
calculation to that of Eq.(\ref{eq:normalization_integral}) provides
\begin{eqnarray}
    &\phantom{=}&\left\langle
      \prod_{\langle i_1,\cdots ,i_K;j_1,\cdots,j_L \rangle}
      \prod_{\alpha=1}^n
      \left\{
        1+\frac{1}{2}
        {\cal D}_{\langle i_1,\cdots ,i_K;j_1,\cdots,j_L \rangle}
        \left(
          S_{i_1}^{\alpha} \cdots S_{i_K}^{\alpha}
            \tau_{j_1}^{\alpha} \cdots \tau_{j_L}^{\alpha}
          -1
        \right)
      \right\}
    \right\rangle_{\cal D} \nonumber \\
    &=&
    {\cal N}^{-1}
    \prod_{i=1}^N
    \left\{
      \oint \frac{d Z_i}{2 \pi i} Z_i^{-(C+1)}
    \right\}
    \prod_{j=1}^M
    \left\{
      \oint \frac{d Y_j}{2 \pi i} Y_j^{-(L+1)}
    \right\} \nonumber \\
    &\phantom{=}&\times
    \prod_{\langle i_1,\cdots ,i_K;j_1,\cdots,j_L \rangle}
    \left[
      1+
      \left(
        Z_{i_1}\cdots Z_{i_K}
        Y_{j_1}\cdots Y_{j_L}
      \right)
      \prod_{\alpha=1}^n
      \frac{1}{2}
      \left(
        1+S_{i_1}^{\alpha}\cdots S_{i_K}^{\alpha}
        \tau_{j_1}^{\alpha}\cdots \tau_{j_L}^{\alpha}
      \right)
    \right]. 
\end{eqnarray}
Using the expansion
\begin{eqnarray}
    &\phantom{=}&\prod_{\alpha=1}^n 
    \left(
      1+S_{i_1}^{\alpha}\cdots S_{i_K}^{\alpha}
      \tau_{j_1}^{\alpha}\cdots \tau_{j_L}^{\alpha}
    \right) \nonumber \\
    &=&
    1
    +\sum_{\alpha=1}^n
    S_{i_1}^{\alpha}\cdots S_{i_K}^{\alpha}
    \tau_{j_1}^{\alpha}\cdots \tau_{j_L}^{\alpha}
    +
    \sum_{\langle \alpha_1,\alpha_2 \rangle}
    (S_{i_1}^{\alpha_1} S_{i_1}^{\alpha_2})
    \cdots
    (S_{i_K}^{\alpha_1} S_{i_K}^{\alpha_2})
    (\tau_{j_1}^{\alpha_1} \tau_{j_1}^{\alpha_2})
    \cdots
    (\tau_{j_L}^{\alpha_1} \tau_{j_L}^{\alpha_2}) \nonumber \\
    &\phantom{=}&+ \cdots +
    \sum_{\langle \alpha_1,\cdots,\alpha_n \rangle}
    (S_{i_1}^{\alpha_1}\cdots S_{i_1}^{\alpha_n})
    \cdots
    (S_{i_K}^{\alpha_1}\cdots S_{i_K}^{\alpha_n})
    (\tau_{j_1}^{\alpha_1}\cdots \tau_{j_1}^{\alpha_n})
    \cdots
    (\tau_{j_L}^{\alpha_1}\cdots \tau_{j_L}^{\alpha_n}) \nonumber \\
    &=&
    \sum_{m=0}^n
    \sum_{\langle \alpha_1,\cdots,\alpha_m \rangle}
    (S_{i_1}^{\alpha_1}\cdots S_{i_1}^{\alpha_m})
    \cdots
    (S_{i_K}^{\alpha_1}\cdots S_{i_K}^{\alpha_m})
    (\tau_{j_1}^{\alpha_1}\cdots \tau_{j_1}^{\alpha_m})
    \cdots
    (\tau_{j_L}^{\alpha_1}\cdots \tau_{j_L}^{\alpha_m}), 
\end{eqnarray}
resulting in
\begin{eqnarray}
    &\phantom{=}&\prod_{\langle i_1,\cdots ,i_K;j_1,\cdots,j_L \rangle}
    \left[
      1+
      \left(
        Z_{i_1}\cdots Z_{i_K}
        Y_{j_1}\cdots Y_{j_L}
      \right)
      \prod_{\alpha=1}^n
      \frac{1}{2}
      \left(
        1+S_{i_1}^{\alpha}\cdots S_{i_K}^{\alpha}
        \tau_{j_1}^{\alpha}\cdots \tau_{j_L}^{\alpha}
      \right)
    \right] \nonumber \\
    &\simeq&
    e^{
    \sum_{\langle i_1,\cdots ,i_K;j_1,\cdots,j_L \rangle}
        Z_{i_1}\cdots Z_{i_K}
        Y_{j_1}\cdots Y_{j_L}
      \prod_{\alpha=1}^n
      \frac{1}{2}
      \left(
        1+S_{i_1}^{\alpha}\cdots S_{i_K}^{\alpha}
        \tau_{j_1}^{\alpha}\cdots \tau_{j_L}^{\alpha}
      \right)
    } \nonumber \\
    &=&
    e^{
    \frac{1}{2^n}
    \sum_{\langle i_1,\cdots ,i_K;j_1,\cdots,j_L \rangle}
        Z_{i_1}\cdots Z_{i_K}
        Y_{j_1}\cdots Y_{j_L}
    \sum_{m=0}^n
    \sum_{\langle \alpha_1,\cdots,\alpha_m \rangle}
    (S_{i_1}^{\alpha_1}\cdots S_{i_1}^{\alpha_m})
    \cdots
    (S_{i_K}^{\alpha_1}\cdots S_{i_K}^{\alpha_m})
    (\tau_{j_1}^{\alpha_1}\cdots \tau_{j_1}^{\alpha_m})
    \cdots
    (\tau_{j_L}^{\alpha_1}\cdots \tau_{j_L}^{\alpha_m})
    }
    \nonumber \\
    &=&
    e^{
      \frac{1}{2^n}
      \left\{
        \sum_{m=0}^n
        \sum_{\langle \alpha_1,\cdots,\alpha_m \rangle}
        \sum_{\langle i_1,\cdots,i_K \rangle}
        \left(
          S_{i_1}^{\alpha_1} \cdots S_{i_1}^{\alpha_m}
          Z_{i_1}
        \right)
        \cdots
        \left(
          S_{i_K}^{\alpha_1} \cdots S_{i_K}^{\alpha_m}
          Z_{i_K}
        \right)
        \sum_{\langle j_1,\cdots,j_L \rangle}
        \left(
          \tau_{j_1}^{\alpha_1} \cdots \tau_{j_1}^{\alpha_m}
          Y_{j_1}
        \right)
        \cdots
        \left(
          \tau_{j_L}^{\alpha_1} \cdots \tau_{j_K}^{\alpha_m}
          Y_{j_L}
        \right)
      \right\}
    }
\nonumber \\
&\simeq&
    e^{
      \frac{1}{2^n}
      \left\{
        \sum_{m=0}^n
        \sum_{\langle \alpha_1,\cdots,\alpha_m \rangle}
        \frac{1}{K!}
        \left(
          \sum_{i=1}^N
          S_i^{\alpha_1} \cdots S_i^{\alpha_m}
          Z_i
        \right)^K
        \frac{1}{L!}
        \left(
          \tau_j^{\alpha_1} \cdots \tau_j^{\alpha_m}
          Y_j
        \right)^L
      \right\}
    }. 
\label{eq:mean_field_representation}
\end{eqnarray}
Using the identities
\begin{eqnarray}
    1&=&\int dq_{\alpha_1,\cdots,\alpha_m}
    \delta
    \left(
      \sum_{i=1}^N S_i^{\alpha_1}\cdots S_i^{\alpha_m}
      Z_i-q_{\alpha_1,\cdots,\alpha_m}
    \right), \nonumber \\
    1&=&\int dr_{\alpha_1,\cdots,\alpha_m}
    \delta
    \left(
      \sum_{j=1}^M \tau_j^{\alpha_1}\cdots \tau_j^{\alpha_m}
      Y_j-r_{\alpha_1,\cdots,\alpha_m}
    \right) \label{eq:delta_conjugate2}
\end{eqnarray}
and going through the same steps as in Eqs. (\ref{eq:delta_conjugate}
- \ref{eq:normalization_extremization}), we arrive at
\begin{eqnarray}
    &\phantom{=}&\prod_{\langle i_1,\cdots ,i_K;j_1,\cdots,j_L \rangle}
    \left[
      1+
      \left(
        Z_{i_1}\cdots Z_{i_K}
        Y_{j_1}\cdots Y_{j_L}
      \right)
      \prod_{\alpha=1}^n
      \frac{1}{2}
      \left(
        1+S_{i_1}^{\alpha}\cdots S_{i_K}^{\alpha}
        \tau_{j_1}^{\alpha}\cdots \tau_{j_L}^{\alpha}
      \right)
    \right] \nonumber \\
    &=&
    \prod_{m=0}^n
    \prod_{\langle \alpha_1,\cdots,\alpha_m \rangle}
    \int dq_{\alpha_1,\cdots,\alpha_m}
    \delta
    \left(
      \sum_{i=1}^N
      S_i^{\alpha_1}\cdots S_i^{\alpha_m} Z_i
      -q_{\alpha1,\cdots,\alpha_m}
    \right) \nonumber \\
    &\phantom{=}&\times
    \int dr_{\alpha_1,\cdots,\alpha_m}
    \delta
    \left(
      \sum_{j=1}^M
      \tau_j^{\alpha_1}\cdots \tau_j^{\alpha_m} Y_j
      -r_{\alpha1,\cdots,\alpha_m}
    \right)
    \exp
    \left(
      \frac{1}{2^n}
      \left\{
        \sum_{m=0}^n
        \sum_{\langle \alpha_1,\cdots,\alpha_m \rangle}
        \frac{q_{\alpha_1,\cdots,\alpha_m}^K}{K!}
        \frac{r_{\alpha_1,\cdots,\alpha_m}^L}{L!}
      \right\}
    \right) \nonumber \\
    &\simeq&
      {\mbox{extr}}_{\{
      \bbox{q},\hat{\bbox{q}},
      \bbox{r},\hat{\bbox{r}}\}
      }
    \Biggl\{
    \exp
    \Biggl[
    \frac{1}{2^n}
    \left\{
      \sum_{m=0}^n
      \sum_{\langle \alpha_1,\cdots,\alpha_m \rangle}
      \frac{q_{\alpha_1,\cdots,\alpha_m}^K}{K!}
      \frac{r_{\alpha_1,\cdots,\alpha_m}^L}{L!}
    \right\}
    \nonumber \\
    &\phantom{=}&-
    \sum_{m=0}^n
    \sum_{\langle \alpha_1,\cdots,\alpha_m \rangle}
    q_{\alpha_1,\cdots,\alpha_m}
    \hat{q}_{\alpha_1,\cdots,\alpha_m}
    -
    \sum_{m=0}^n
    \sum_{\langle \alpha_1,\cdots,\alpha_m \rangle}
    r_{\alpha_1,\cdots,\alpha_m}
    \hat{r}_{\alpha_1,\cdots,\alpha_m} \nonumber \\
    &\phantom{=}&+
    \sum_{m=0}^n
    \sum_{\langle \alpha_1,\cdots,\alpha_m \rangle}
    \hat{q}_{\alpha_1,\cdots,\alpha_m}
    \sum_{i=1}^N
    S_i^{\alpha_1}\cdots S_i^{\alpha_m}Z_i
    +
    \sum_{m=0}^n
    \sum_{\langle \alpha_1,\cdots,\alpha_m \rangle}
    \hat{r}_{\alpha_1,\cdots,\alpha_m}
    \sum_{j=1}^M
    \tau_j^{\alpha_1}\cdots \tau_j^{\alpha_m}Y_j
    \Biggr]
    \Biggr\}. \label{eq:extremized_product}
\end{eqnarray}

In order to proceed further, one has to make an assumption about the
order parameter symmetry. We adopt here the replica symmetric ansatz
for the order parameters $q$, $r$, $\hat{q}$ and $\hat{r}$. This
implies that the order parameters do not depend on the explicit
indices but only on their number. It is therefore convenient to
represent them as moments of random variables defined over the
interval $[-1,1]$
\begin{eqnarray}
    q_{\alpha_1,\cdots,\alpha_l}
    &=&
    q \int dx \ \pi (x)  \ x^l, \quad
    r_{\alpha_1,\cdots,\alpha_l}
    =
    r \int dy \ \rho (y)  \ y^l, \nonumber \\
    \hat{q}_{\alpha_1,\cdots,\alpha_l}
    &=&
    \hat{q} \int d\hat{x} \ \hat{\pi} (\hat{x})  \ \hat{x}^l, \quad
    \hat{r}_{\alpha_1,\cdots,\alpha_l}
    =
    \hat{r} \int d\hat{y} \ \hat{\rho} (\hat{y})  \ \hat{y}^l, 
    \label{eq:rs_ansatz}
\end{eqnarray}
Then, each term in Eq.(\ref{eq:extremized_product}) takes the form
\begin{eqnarray}
    \sum_{m=0}^n
    \sum_{\langle \alpha_1,\cdots,\alpha_m \rangle}
    \frac{q_{\alpha_1,\cdots,\alpha_m}^K}{K!}
    \frac{r_{\alpha_1,\cdots,\alpha_m}^L}{L!}
    &=&
    \frac{q^K}{K!}\frac{r^L}{L!}
    \sum_{m=0}^n
    \left( n \atop m \right) 
    \int \prod_{k=1}^K \ dx_k \pi(x_k) x_k^m
    \int \prod_{l=1}^L \ dy_l \rho(y_l) y_l^m \nonumber \\
    &=&
    \frac{q^K}{K!}\frac{r^L}{L!}
    \int \prod_{k=1}^K dx_k \ \pi(x_k)
    \int \prod_{l=1}^L dy_l \ \rho(y_l)
    \left(
      1+\prod_{k=1}^K x_k \prod_{l=1}^L y_l
    \right)^n \label{app1:eqn33} \\
    \sum_{m=0}^n
    \sum_{\langle \alpha_1,\cdots,\alpha_m \rangle}
    q_{\alpha_1,\cdots,\alpha_m}
    \hat{q}_{\alpha_1,\cdots,\alpha_m}
    &=&
    q \hat{q}
    \sum_{m=0}^n \left( n \atop m \right)
    \int dx \ d \hat{x} \
    \pi(x)  \ \hat{\pi}(\hat{x})
    x^m  \ \hat{x}^m \nonumber \\
    &=&
    q \hat{q}
    \int dx \ d \hat{x} \ \pi(x)  \ \hat{\pi}(\hat{x})
    \left(
      1+x \hat{x}
    \right)^n \label{app1:eqn34} \\
    \sum_{m=0}^n
    \sum_{\langle \alpha_1,\cdots,\alpha_m \rangle}
    \hat{q}_{\alpha_1,\cdots,\alpha_m}
    \sum_{i=1}^N
    S_i^{\alpha_1}\cdots S_i^{\alpha_m}Z_i
    &=&
    \hat{q} \sum_{i=1}^N Z_i
    \int d \hat{x} \ \hat{\pi}(\hat{x})
    \sum_{m=0}^n
    \hat{x}^m \sum_{\langle \alpha_1,\cdots,\alpha_m \rangle}
    S_i^{\alpha_1}\cdots S_i^{\alpha_m} \nonumber \\
    &=&
    \hat{q} \sum_{i=1}^N Z_i
    \int d \hat{x} \ \hat{\pi}(\hat{x})
    \prod_{\alpha=1}^n
    \left(
      1+S_i^{\alpha} \hat{x}
    \right) \ . \label{app1:eqn35}
\end{eqnarray}

Substituting these into (\ref{eq:extremized_product}), one obtains
\begin{eqnarray}
    &\phantom{=}&\langle
    Z(\bbox{\xi},\bbox{\zeta},{\cal D})^n
    \rangle_{\bbox{\xi},\bbox{\zeta},{\cal D}} \nonumber \\
    &=&
    \sum_{\bbox{S}^1 \cdots \bbox{S}^n}
    \sum_{\bbox{\tau}^1 \cdots \bbox{\tau}^n}
    \prod_{i=1}^N
    \left \langle 
      \exp \left(
        \xi  F_s \sum_{\alpha=1}^n S_i^{\alpha}
      \right)
    \right \rangle_{\xi} \times
    \prod_{j=1}^M
    \left \langle 
      \exp  \left(
        \zeta F_n \sum_{\alpha=1}^n \tau_j^{\alpha}
      \right)
    \right \rangle_\zeta \nonumber \\
    &\phantom{=}&\times
    {\cal N}^{-1}
    \prod_{i=1}^N
    \left\{
      \oint \frac{d Z_i}{2 \pi i} Z_i^{-(C+1)}
    \right\}
    \prod_{j=1}^M
    \left\{
      \oint \frac{d Y_j}{2 \pi i} Y_j^{-(L+1)}
    \right\} \nonumber \\
    &\phantom{=}&\times
      {\mbox{extr}}_{\{ \pi,\hat{\pi},\rho,\hat{\rho}\}}
    \Biggl\{
    \exp
    \Biggl[
    \frac{1}{2^n}
    \left\{
      \frac{q^K}{K!}
      \frac{r^L}{L!}
      \int \prod_{l=1}^K dx_l \ \pi (x_l)
      \int \prod_{l=1}^L dy_l \ \rho (y_l)
      \left(
        1+\prod_{l=1}^K x_l \prod_{l=1}^L y_l
      \right)^n
    \right\} \nonumber \\
    &\phantom{=}&
    -q \hat{q} \int dx \ d \hat{x} \
    \pi (x) \ \hat{\pi}(\hat{x})
    (1+x \hat{x})^n
    -r \hat{r} \int dy \ d \hat{y} \
    \rho (y) \ \hat{\rho}(\hat{y})
    (1+y \hat{y})^n \nonumber \\
    &\phantom{=}&
\left . \left .   +\hat{q}\sum_{i=1}^N Z_i
    \int d \hat{x} \ \hat{\pi}(\hat{x})
    \prod_{\alpha=1}^n
    (1+S_i^{\alpha} \hat{x})
   +\hat{r} \sum_{j=1}^M Y_j
    \int d \hat{y} \ \hat{\rho}(\hat{y})
    \prod_{\alpha=1}^n
    (1+\tau_j^{\alpha} \hat{y}) \right ] \right \}. \label{eq:Zn}
\end{eqnarray}
The term involving the spin variables $S$ is easily evaluated using
the residue theorem
\begin{eqnarray}
    &\phantom{=}&\sum_{\bbox{S}^1 \cdots \bbox{S}^n}
    \prod_{i=1}^N
    \left \langle 
      \exp      \left(
        \xi F_s \sum_{\alpha=1}^n S_i^{\alpha}
      \right)
    \right \rangle_\xi
    \prod_{i=1}^N
    \left\{
      \oint \frac{d Z_i}{2 \pi i} Z_i^{-(C+1)}
    \right\} 
    \times \exp    \left[
      \hat{q}\sum_{i=1}^N Z_i
      \int d \hat{x} \ \hat{\pi}(\hat{x})
      \prod_{\alpha=1}^n
      (1+S_i^{\alpha} \hat{x})
    \right] \nonumber \\
    &=&
    \left(
      \frac{\hat{q}^C}{C!}
      \int \prod_{l=1}^C 
      d \hat{x}_l \ \hat{\pi}(\hat{x}_l)
      \left\langle
        \prod_{\alpha=1}^n
        \left\{
          e^{\xi F_s} \prod_{l=1}^C(1+\hat{x}_l)
          +
          e^{-\xi F_s} \prod_{l=1}^C(1-\hat{x}_l)
        \right\}
      \right\rangle_{\xi}
    \right)^N,  \label{eq:sum_of_S}
\end{eqnarray}
and similarly for the term involving the variables $\tau$.
Substituting these into Eq. (\ref{eq:Zn}), one obtains the $n$-th
moment of partition function
\begin{eqnarray}
    &\phantom{=}& 
\left \langle {\cal Z}(\bbox{\xi},\bbox{\zeta},{\cal D})^n
    \right \rangle_{\xi,\zeta,{\cal D}} \nonumber \\
    &=&
     {\mbox{extr}}_{\{ \pi,\hat{\pi},\rho,\hat{\rho} \}}
    \Biggl\{
    \exp
    \Biggl[
    -NC
    \left\{
      \int dx \ d \hat{x}  \ \pi (x) \ \hat{\pi}(\hat{x}) \ln 
      (1+x \hat{x})^n-1
    \right\} 
\nonumber \\
   &\phantom{=}&
    -ML
    \left\{
      \int dy  \ d \hat{y} \  \rho(y) \ 
      \hat{\rho}(\hat{y}) \ln (1+y \hat{y})^n-1
    \right\} \nonumber \\
    &\phantom{\times}&
    +\frac{1}{2^n}
    \left\{
      \frac{NC}{K}
      \int \left[ \prod_{k=1}^K
        dx_k \  \pi (x_k) \right]
      \left[ \prod_{l=1}^L dy_l \rho(y_l) \right]
      \ln \left[ 1+\prod_{k=1}^K x_k \prod_{l=1}^L y_l
      \right]^n-1
    \right\}
    \Biggr] \nonumber \\
    &\phantom{=}& \times
    \left(
      \int 
      \left[ \prod_{k=1}^C d \hat{x}_k 
        \hat{\pi}(\hat{x}_k) 
      \right]
      \left \langle
        \left(
          \left[
            e^{F_s \xi} \prod_{k=1}^C (1+\hat{x}_k)
            +e^{-F_s \xi}  \prod_{k=1}^C (1-\hat{x}_k)
          \right]
        \right)^n
      \right \rangle _{\xi}
    \right)^N \nonumber \\
    &\phantom{=}&\times
    \left(
      \int 
      \left[ \prod_{l=1}^L d \hat{y}_l 
        \hat{\rho}(\hat{y}_l) 
      \right]
      \left \langle 
        \left(
          \left[
            e^{F_n \zeta} \prod_{l=1}^L (1+\hat{y}_l)
            +e^{-F_n \zeta}  \prod_{l=1}^L (1-\hat{y}_l)
          \right]
        \right)^n
      \right \rangle _{\zeta}
    \right)^M
    \Biggr\}. \label{eq:extremized_Zn} 
\end{eqnarray} 
Finally, in the limit $n \to 0$ one obtains
\begin{eqnarray}
    &\phantom{=}& \frac{1}{N} \langle \ln
    {\cal Z}(\bbox{\xi},\bbox{\zeta},{\cal D}) \rangle
    _{\xi,\zeta,{\cal D}} = 
\lim_{n \to 0} \frac{\left \langle {\cal Z}(\bbox{\xi},\bbox{\zeta},{\cal D})^n
\right \rangle_{\xi,\zeta,{\cal D}}-1}{nN}    \nonumber \\
    &=& 
{\mbox{extr}}_{\{ \pi,\hat{\pi},\rho,\hat{\rho} \}}
\left \{ -\frac{C}{K}\ln 2
    -C \int dx \ d \hat{x}  \ \pi (x) \hat{\pi} \ (\hat{x}) \ln 
    (1+x \hat{x})
    - \frac{CL}{K} \int dy  \ d \hat{y}  \ \rho(y) \ 
    \hat{\rho}(\hat{y}) \ln (1+y \hat{y})  \right . 
\nonumber \\
    &\phantom{=}&+
      \frac{C}{K}\int \left[ \prod_{k=1}^K
      dx_k \pi (x_k) \right]
    \left[ \prod_{l=1}^L dy_l \rho(y_l) \right]
    \ln \left[ 1+\prod_{k=1}^K x_k \prod_{l=1}^L y_l \right] \nonumber \\
    &\phantom{=}&+ \int \left[ \prod_{k=1}^C d \hat{x}_k 
      \hat{\pi}(\hat{x}_k) \right]
    \left \langle \ln \left[
        e^{F_s \xi} \prod_{k=1}^C (1+\hat{x}_k)
        +e^{-F_s \xi}  \prod_{k=1}^C (1-\hat{x}_k)
      \right]
    \right \rangle _{\xi} \nonumber \\
 &\phantom{=}& \left .  
+\frac{C}{K} \int \left[ \prod_{l=1}^L d \hat{y}_l 
      \hat{\rho}(\hat{y}_l) \right]
    \left \langle \ln \left[
        e^{F_n \zeta} \prod_{l=1}^L (1+\hat{y}_l)
        +e^{-F_n \zeta}  \prod_{l=1}^L (1-\hat{y}_l)
      \right]
    \right \rangle _{\zeta} \right \}. 
\label{eq:logZ} 
\end{eqnarray} 

\section{Evaluation of the Magnetisation}
Here, we derive explicitly Eqs.(\ref{eq:magnetization}) and
(\ref{eq:physical_field}).  After using the gauge transformation $S_i \to
\xi_i S_i$, the magnetisation can be written as
\begin{equation}
m=\frac{1}{N} \sum_{i=1}^N \left \langle 
\mbox{sign} (m_i)
\right \rangle_{\mxi,\mzeta,{\cal D}} \ ,
\label{eq:mag_app}
\end{equation}
introducing the  notation 
$m_i=\left \langle S_i \right \rangle_{\beta \to \infty}$
(gauged average). 

For an arbitrary natural number $p$, one can compute 
$p$-th moment of $m_i$ 
\begin{equation}
\left \langle 
{m_i}^p
\right \rangle_{\mxi,\mzeta,{\cal D}}= 
\lim_{n \to 0} \lim_{\beta \to \infty}
\left \langle \sum_{\{\mS^1,\mtau^1\}, 
\ldots,\{\mS^n,\mtau^n\}}
S_i^1 \cdot S_i^2 \cdot \ldots \cdot S_i^p e^{-\beta \sum_{a=1}^n {\cal H}_a}
\right\rangle_{\mxi,\zeta,{\cal D}}, 
\label{eq:pthmoment_definition}
\end{equation}
where ${\cal H}_a$ denotes the gauged Hamiltonian of the $a$-th
replica.  Decoupling the dynamical variables and introducing auxiliary
functions $\pi(\cdot)$, $\hat{\pi}(\cdot)$, $\rho(\cdot)$ and
$\hat{\rho}(\cdot)$, of a similar form to Eq. (\ref{eq:rs_ansatz}),
one obtains
\begin{equation}
\left \langle 
{m_i}^p
\right \rangle_{\mxi,\mzeta,{\cal D}}=\int
\prod_{l=1}^C d \hat{x}_l \ \hat{\pi}(\hat{x}_l) 
\left \langle 
\tanh^p\left( F_s \xi + \sum_{k=1}^C \tanh^{-1} \hat{x}_k \right )
\right \rangle_\xi \ ,
\label{eq:pthmoment}
\end{equation}
using the saddle point solution of $\hat{\pi}(\cdot)$. 

Employing the identity
\begin{equation}
\mbox{sign} (x)=-1+2 
\lim_{n \to \infty} \sum_{m=0}^n \left( 2 n \atop m \right)
\left ( \frac{1+x}{2} \right )^{2n-m}
\left ( \frac{1-x}{2} \right )^{m}
\label{eq:sign_x}
\end{equation}
which holds for any arbitrary 
real number $x \in [-1,1]$ and Eqs.(\ref{eq:pthmoment})
and (\ref{eq:sign_x}) one obtains
\begin{eqnarray}
\left \langle 
\mbox{sign} (m_i)
\right \rangle_{\mxi,\mzeta,{\cal D}}
&=& -1+2 \int d z \ \phi(z)  \
\lim_{n \to \infty} \sum_{m=0}^n \left( 2 n \atop m \right)
\left ( \frac{1+z}{2} \right )^{2n-m}
\left ( \frac{1-z}{2} \right )^{m} \cr
&=& \int d z \ \phi(z) \ \mbox{sign} (z), 
\label{eq:sign_m}
\end{eqnarray}
where we introduced a new notation for the  distribution 
\begin{equation}
\phi(z)=\int
\prod_{l=1}^C d \hat{x}_l \ \hat{\pi}(\hat{x}_l) 
\left \langle \delta (z-F_s \xi - \sum_{k=1}^C \tanh^{-1} \hat{x}_k)
\right \rangle_\xi \ , 
\label{eq:phi}
\end{equation}
thus reproducing Eqs.(\ref{eq:magnetization}) and (\ref{eq:physical_field}).

\begin{figure}
\begin{center} \leavevmode
\epsfig{file=free_energy_all.eps,angle=0,width=150mm}
\end{center}
\caption{Left hand figures show a schematic representation of the
free energy landscape while figures on the right show the
ferromagnetic, sub-optimal ferromagnetic and paramagnetic solutions as
functions of the noise rate $p$; thick and thin lines denote stable
solutions of lower and higher free energies respectively, dashed lines
correspond to unstable solutions.
(a) $K \ge 3$ or $L \ge 3$, $K>1$; the solid line in the horizontal
axis represents the phase where the ferromagnetic solution (F, $m=1$)
is thermodynamically dominant, while the paramagnetic solution (P,
$m=0$) becomes dominant for the other phase (dashed line).  The
critical noise $p_c$ denotes Shannon's channel capacity.
(b) $K = 2$ and $L = 2$; the ferromagnetic solution and its mirror
image are the only minima of the free energy over a relatively small
noise level (the solid line in the horizontal).  The critical
point,due to dynamical considerations, is the spinodal point $p_s$
where sub-optimal ferromagnetic solutions (F', $m<1$) emerge.  The
thermodynamic transition point $p_3$, at which the ferromagnetic
solution loses its dominance, is below the maximum noise level given
by the channel capacity, which implies that these codes do not
saturate Shannon's bound even if optimally decoded.
(c) $K=1$; the solid line in the horizontal axis represents the range
of noise levels where the ferromagnetic state (F) is the only minimum
of the free energy. The sub-optimal ferromagnetic state (F') appears
in the region represented by the dashed line. The spinodal point
$p_s$, where F' solution first appears, provides the highest noise
value in which convergence to the ferromagnetic solution is
guaranteed. For higher noise levels, the system becomes bistable and
an additional unstable solution for the saddle point equations
necessarily appears. A thermodynamical transition occurs at the noise
level $p_1$ where the state F' becomes dominant.}

\label{landscape_both}
\end{figure}
\newpage 



\newpage

\newpage

\begin{table}
\vspace*{1.5cm}
 \caption{ Comparison between the maximal tolerable noise level for
codes based on randomly and systematically structured matrices in the
case of $K=L=2$; decoding is carried out using BP/TAP and the
transmission channel used is the BSC.  The performance of both matrix
structures is highly similar. }
\vspace*{1.5cm}
 \begin{tabular}{c|ddddddd} 
 Rate $R=K/C$ & 0.6666 & 0.5 & 0.4 & 0.3333 & 0.2857 & 0.2 & 0.1 \\
 \tableline
 \tableline
 Systematic Matrix& 0.0527 & 0.0934 & 0.1222 & 0.1416 & 0.1598 & 
0.1927 & 0.2476 \\
 & $\pm$0.0016 & $\pm$0.0019 & $\pm$0.0012 & $\pm$0.0016 & $\pm$0.0007 & $\pm$0.0016 & $\pm$0.0010 \\
 \tableline
 Random Matrix& 0.0528 & 0.0930 & 0.1206 & 0.1439 & 
0.1599 & 0.1931 & 0.2477 \\
 & $\pm$0.0009 & $\pm$0.0019 & $\pm$0.0010 & $\pm$0.0017 & $\pm$0.0010 & $\pm$0.0014 & $\pm$0.0014 \\
 \end{tabular}
 \label{performance}
\end{table}

\end{document}